# Hybrid Learning: A Combination of Self-Supervised and Supervised Learning for Joint MRI Reconstruction and Denoising in Low-Field MRI

Haoyang Pei[1,2,3], Nikola Janjušević[1,2], Renqing Luo[1,2,3], Ding Xia[4], Xiang Xu[4], William Moore[1,2], Yao Wang[3], Hersh Chandarana[1,2], Li Feng[1,2]

1 Bernard and Irene Schwartz Center for Biomedical Imaging, Department of Radiology, New York University Grossman School of Medicine, New York, NY, USA.

2 Center for Advanced Imaging Innovation and Research (CAI$^2$R), Department of Radiology, New York University Grossman School of Medicine, New York, NY, USA.

3 Department of Electrical and Computer Engineering, NYU Tandon School of Engineering, New York, NY, USA

4 Biomedical Engineering and Imaging Institute and Department of Radiology, Icahn School of Medicine at Mount Sinai, New York, NY, United States

**Word Count:** 5270

**Running Header:** Hybrid Learning for MRI Reconstruction

**Grant Support:** NIH R01EB030549, R01EB031083, R21EB032917, and P41EB017183

Address correspondence to:

Li Feng, PhD

Center for Advanced Imaging Innovation and Research (CAI$^2$R)

New York University Grossman School of Medicine

227 E 30th St

New York, NY, USA 10016

Email: Li.Feng@nyulangone.org

## Abstract

Deep learning has demonstrated strong potential for MRI reconstruction. However, conventional supervised learning requires high-quality, high-SNR references for network training, which are often difficult or impossible to obtain in different scenarios, particularly in low-field MRI. Self-supervised learning provides an alternative by removing the need for training references, but its reconstruction performance can degrade when the baseline SNR is low. To address these limitations, we propose hybrid learning, a two-stage training framework that integrates self-supervised and supervised learning for joint MRI reconstruction and denoising when only low-SNR training references are available. Hybrid learning is implemented in two sequential stages. In the first stage, self-supervised learning is applied to fully sampled low-SNR data to generate higher-quality pseudo-references. In the second stage, these pseudo-references are used as targets for supervised learning to reconstruct and denoise undersampled noisy data. The proposed technique was evaluated in multiple experiments involving simulated and real low-field MRI in the lung and brain at different field strengths. Hybrid learning consistently improved image quality over both standard self-supervised learning and supervised learning with noisy training references at different acceleration rates, noise levels, and field strengths, achieving higher SSIM and lower NMSE. The hybrid learning approach is effective for both Cartesian and non-Cartesian acquisitions. Hybrid learning provides an effective solution for training deep MRI reconstruction models in the absence of high-SNR references. By improving image quality in low-SNR settings, particularly for low-field MRI, it holds promise for broader clinical adoption of deep learning–based reconstruction methods.

## Introduction

Rapid MRI techniques play an essential role in clinical imaging by enabling faster scans without compromising diagnostic quality. Traditional acceleration methods, such as parallel imaging, compressed sensing, and their variants, have demonstrated significant value in accelerating MRI acquisition across a wide range of clinical applications (1-10). In recent years, deep learning-based MRI reconstruction methods have shown remarkable promise for fast MRI, enabling higher acceleration rates and improved image quality that were previously considered unattainable (11-32). To date, state-of-the-art deep learning-based MRI reconstruction techniques often rely on supervised learning, which trains neural networks using fully (or sufficiently) sampled, high-quality reference images (12-20). However, this dependence on high-quality references poses a major challenge in scenarios where such high-quality datasets are difficult or impossible to obtain. For example, at lower field strengths (e.g., 0.55T or below), acquiring fully sampled reference images with high signal-to-noise ratio (SNR) can be time-consuming or even in practical.

Self-supervised learning methods have recently been developed to address this limitation by enabling MRI reconstruction without requiring fully sampled references (21-32). While these approaches can achieve reconstruction quality comparable to supervised learning within certain acceleration ranges, they tend to degrade at higher accelerations, particularly when the baseline SNR is low (33). These challenges are especially relevant in low-field MRI, where the intrinsically low SNR further complicates the reconstruction problem. As a result, existing supervised and self-supervised approaches may face important limitations in reconstructing accelerated data in low-field MRI applications.

To address this need, this work proposes a deep learning-based MRI reconstruction strategy, called hybrid learning, for joint denoising and reconstruction when only fully sampled low-SNR training references are available. Hybrid learning employs a two-stage framework that effectively combines the strengths of self-supervised and supervised learning while mitigating their respective limitations. In the first stage, self-supervised learning is applied to fully sampled low-SNR reference data to reconstruct higher-quality pseudo-reference images. In the next stage, these pseudo-references generated from the first stage then serve as training targets for supervised learning to reconstruct undersampled low-SNR data. Once training is complete, the second-stage network can be directly applied to reconstruct new undersampled low-SNR datasets.

The performance of the proposed hybrid learning technique was evaluated for joint MRI reconstruction and denoising using both simulated and real low-field noisy data. The overall hypothesis is that by leveraging self-supervised learning to generate intermediate high-quality pseudo-references, hybrid learning reduces the reliance on fully sampled, high-quality references for supervised training, while achieving superior reconstruction performance compared to either conventional supervised or self-supervised learning alone.

## Methods

In this section, we review the fundamentals of deep learning-based MRI reconstruction, outline the supervised and self-supervised learning strategies adopted in this work, and describe how hybrid learning is derived by combining these two approaches. Subsequently, the proposed hybrid framework is evaluated through several experiments in both simulated noisy data and real low-field acquisitions.

### MRI Reconstruction Using Unrolled Deep Learning Architectures

Equation 1 below defines the optimization problem for constrained MRI reconstruction that can be applied to undersampled Cartesian or non-Cartesian data.

$$\hat{x} = \underset{x}{\operatorname{argmin}} \frac{1}{2} ||\sqrt{W}(Ex - y)||_2^2 + \lambda \mathcal{R}(x) \tag{1}$$

Here, $E = FC$ represents the multi-coil encoding operator, which integrates the coil sensitivity maps ($C$) and the Fourier transform ($F$). For Cartesian sampling, $F$ represents a standard partial Fourier sampling at sampling mask $M$. Note that, to achieve a unified formulation for Cartesian and non-Cartesian MRI sampling, the sampling mask $M$ is incorporated into the definition of the FFT operator. For non-Cartesian sampling, $F$ represents the non-uniform fast Fourier transform (NUFFT) incorporating the underlying sampling trajectory and the encoding operator becomes $E = \sqrt{W}FC$, where a density compensation matrix $W$ is often applied to account for the non-uniform k-space sampling density in non-Cartesian trajectories. This stabilizes convergence and improves the reconstructed performance during iterative reconstruction (34). $W$ is split into two square-root terms to ensure that the multi-coil encoding operator and its Hermitian transpose are adjoint. In Cartesian sampling, $W$ can be simply treated as the identity matrix ($W = I$). $y$ denotes the acquired multi-coil k-space data, and $x$ represents the coil-combined image to be reconstructed. A regularization term $\mathcal{R}(\cdot)$ is included in image reconstruction with a corresponding parameter $\lambda$.

Equation 1 can be solved using different algorithms, such as the gradient descent approach. In a deep learning framework, this can be implemented with an unrolled network, where the regularization term is learned through a series of small convolutional neural networks (CNNs) to iteratively solve the optimization problem, as expressed in Equation 2 (15).

$$x^{i+1} = x^i - \mu^i C^H F^H \sqrt{W}\left(\sqrt{W}FCx^i - \sqrt{W}y\right) - \mathrm{CNN}\left(x^i\right) \quad (2)$$

Here, $\mu^i$ is a learnable parameter for the $i^{th}$ unrolled block in the neural network. As described in (15), incorporating $C$ to both sides of Equation 2 gives the corresponding update equation in the multi-coil image:

$$x_{mc}^{i+1} = x_{mc}^i - \mu^i\left(F^H W F x_{mc}^i - x_{mc}^0\right) - C\left[\mathrm{CNN}\left(C^H x_{mc}^i\right)\right] \quad (3)$$

where $x_{mc}^0$ denotes the initial multi-coil undersampled image, and $\mu^i$ is a learnable parameter for each update step $i$.

Based on equation 3, an unrolled network architecture can be constructed for both non-Cartesian and Cartesian MRI reconstructions as shown in **Supporting Information Figure S1 and S2**. The reconstruction module employs multiple small U-Nets (35) that can model iterative gradient descent updates in equation 3, where CNNs can be used to learn the regularization function. In addition, the unrolled network can also incorporate an additional module for joint coil sensitivity estimation. As described in (15), this module can be implemented using U-Net to estimate coil sensitivity maps from the k-space center, and it can be integrated into the reconstruction process to improve image quality.

During reconstruction, the model takes the initial multi-coil image ($x_{mc}^{0} = F^{H}Wy$) as input. The coil sensitivity estimation module first estimates the coil sensitivity maps, and the reconstruction module applies the iterative gradient descent updates in equation 3 to the input multi-coil image $x_{mc}^{0}$ for generating the reconstructed multi-coil image $x_{mc}^{I}$. Finally, the estimated coil sensitivity maps are used to combine $x_{mc}^{I}$ into the final output coil-combined image $\hat{x}$.

**Supervised and Self-Supervised Learning for MRI Reconstruction**

To optimize the reconstruction network described above, an appropriate loss function needs to be constructed. Depending on how the loss is defined, deep learning-based MRI reconstruction methods can be broadly categorized into supervised and self-supervised learning.

In supervised learning, a reconstruction network $f_{\theta}$ is optimized by minimizing the expected difference between images reconstructed from undersampled data $\hat{x} = f_{\theta}(F^{H}Wy)$ and fully sampled (or sufficiently sampled without noted artifacts), high-quality reference images $x_{\text{ref}}$ using a predefined loss function $\mathcal{L}_{SL}$.

$$\theta = \underset{\theta}{\operatorname{argmin}}(\mathbb{E}[\mathcal{L}_{SL}(\hat{x}, x_{\text{ref}})]) \tag{4}$$

In practice, this expectation is approximated by averaging the loss across all samples in the training datasets. Once training is complete, the network can be directly applied to new undersampled data for image reconstruction. This approach is widely used and has become the benchmark in many studies (12-20). However, the need for high-quality references poses a major challenge in scenarios where such datasets are difficult to obtain, such as at low-field strengths.

Self-supervised learning (SSL) has emerged as a promising alternative to overcome this limitation. SSL enables MRI reconstruction from undersampled data without the need for fully sampled reference images (21-32). A widely adopted training approach in self-supervised learning-based image reconstruction employs a data consistency (DC) loss or a model consistency loss to ensure that the reconstructed images remain consistent with the acquired undersampled measurements. Additional priors or specifically designed regularizations, such as measurement splitting and equivariant transformations, can be incorporated to further improve reconstruction quality. A widely recognized self-supervised MRI reconstruction method is SSDU (self-supervised learning via data undersampling) proposed by Yaman et al (21). In SSDU, each undersampled k-space dataset is divided into two disjoint subsets (Set A and B, referred to as $y_A$ and $y_B$): one set (Set A) for training a network and the other set (Set B) for ensuring DC during network training. The reconstruction network $f_\theta$ can be optimized by this self-supervised loss $\mathcal{L}_{SSL}$ as:

$$\theta = \underset{\theta}{\operatorname{argmin}}(\mathbb{E}[\,\mathcal{L}_{SSL}(F_B^H W_B F_B C f_\theta(F_A^H W_A y_A), F_B^H W_B y_B)]) \quad (5)$$

Here, $F_A$ and $F_B$ represent two FFT/NUFFT operators for Set A and Set B, incorporating corresponding sampling mask/trajectories, respectively. Once trained, the network can be applied to reconstruct new undersampled MRI data without data splitting. In addition to image reconstruction, this training strategy can also effectively act as a denoising scheme because the noise in Set A and Set B is independent, while the underlying anatomical structure remains the same. As a result, the model learns to reconstruct clean images by minimizing the differences between pairs of noisy observations without the need for ground-truth clean images, following the Noise2Noise model (36). More recently, the SSDU technique has also been extended to different variants to achieve improved reconstruction performance, as described in corresponding literatures (23, 29, 37, 38).

In addition to measurement splitting, several other self-supervised learning strategies have also been proposed. For example, Liu et al. proposed a self-supervised learning approach based on iterative data refinement, which employs multi-stage self-supervised training to progressively improve the quality of the training data and thereby reduce the bias introduced by the self-supervised data setting (32). Chen et al. developed Equivariant Imaging, a self-supervised reconstruction method that incorporates both the DC loss and an invariant set consistency loss by leveraging natural signal equivariances (24, 31). However, although self-supervised reconstruction techniques can achieve performance comparable to supervised methods at moderate acceleration rates, their effectiveness in reconstructing undersampled low-SNR data remains unclear (33).

**Hybrid Learning: A Combination of Self-Supervised and Supervised Learning**

To address the limitations of both supervised and self-supervised learning, we propose a hybrid learning strategy that combines the two approaches within a two-stage training framework to enable joint MRI reconstruction and denoising when only fully sampled low-SNR training references are available.

We assume access to a fully sampled, low-SNR k-space database with N datasets, denoted as $D^{(1)} = \left\{y_j^{(1)}\right\}_{j=1}^{N}$. From $D^{(1)}$, an undersampled, low-SNR k-space database $D^{(2)} = \left\{y_j^{(2)}\right\}_{j=1}^{N}$, can be generated by retrospectively undersampling each k-space data in $D^{(1)}$ according to a predefined acceleration rate, achieved by masking or discarding a subset of k-space samples. The goal of hybrid learning is to develop a two-stage reconstruction model that recovers high-quality images from $y^{(2)}$ through joint reconstruction and denoising.

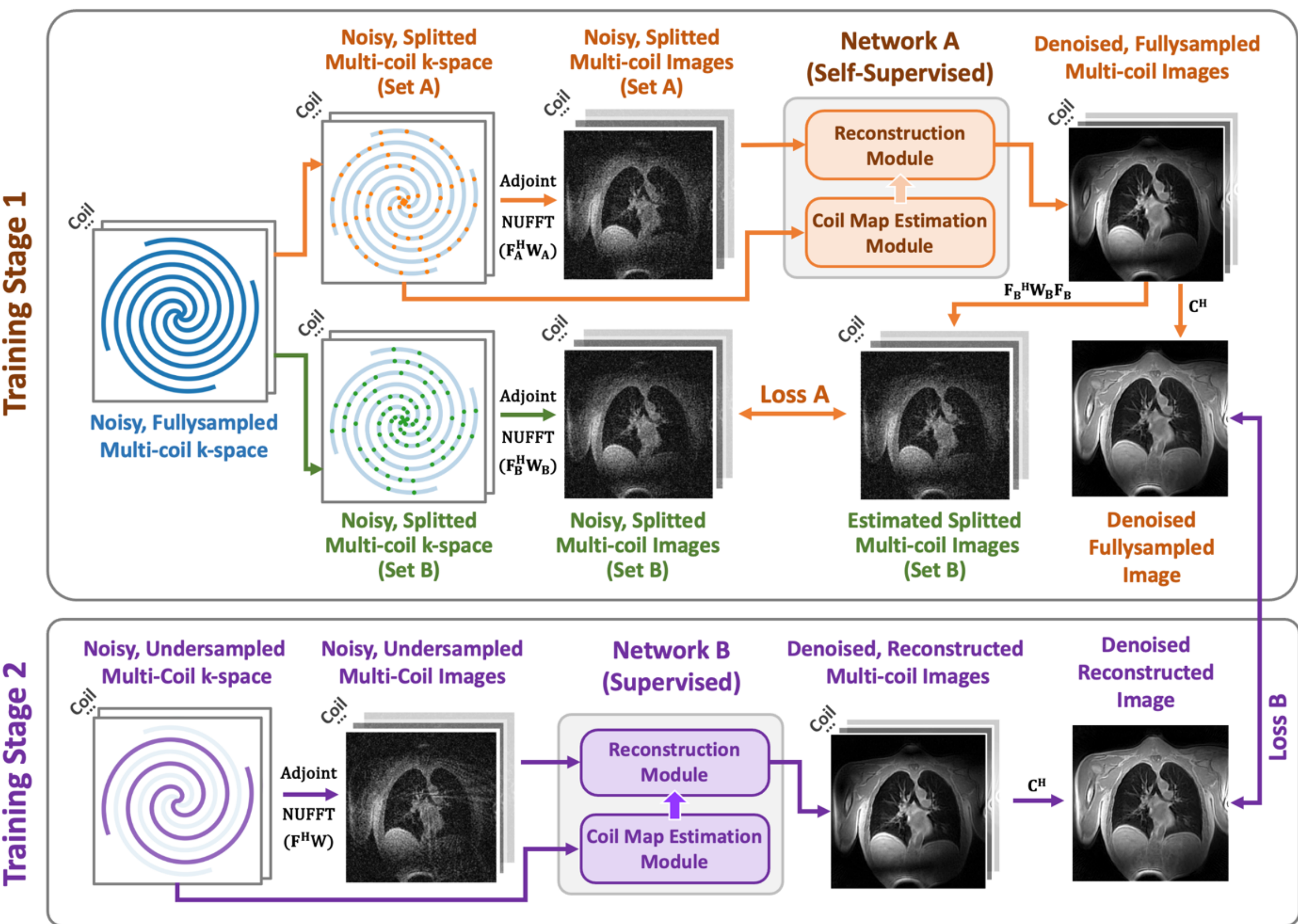

**Figure 1:** The hybrid learning pipeline for non-Cartesian (spiral) sampling applications (e.g., breath-hold Spiral MRI of the Lung at 0.55T). In the first stage, Network A is trained for image denoising using fully sampled, low-SNR spiral datasets. Self-supervised learning is employed in this step using an approach adapted from the SSDU technique. Once training is complete, inference can be performed directly on fully sampled, low-SNR spiral datasets without k-space splitting to generate denoised coil-combined images as high-quality pseudo-references. In the second stage, Network B is trained on the same datasets used in the first stage for joint reconstruction and denoising using the high-quality pseudo-references as training targets. Once training is completed, Network B can be directly applied to new undersampled, low-SNR datasets for joint denoising and reconstruction.

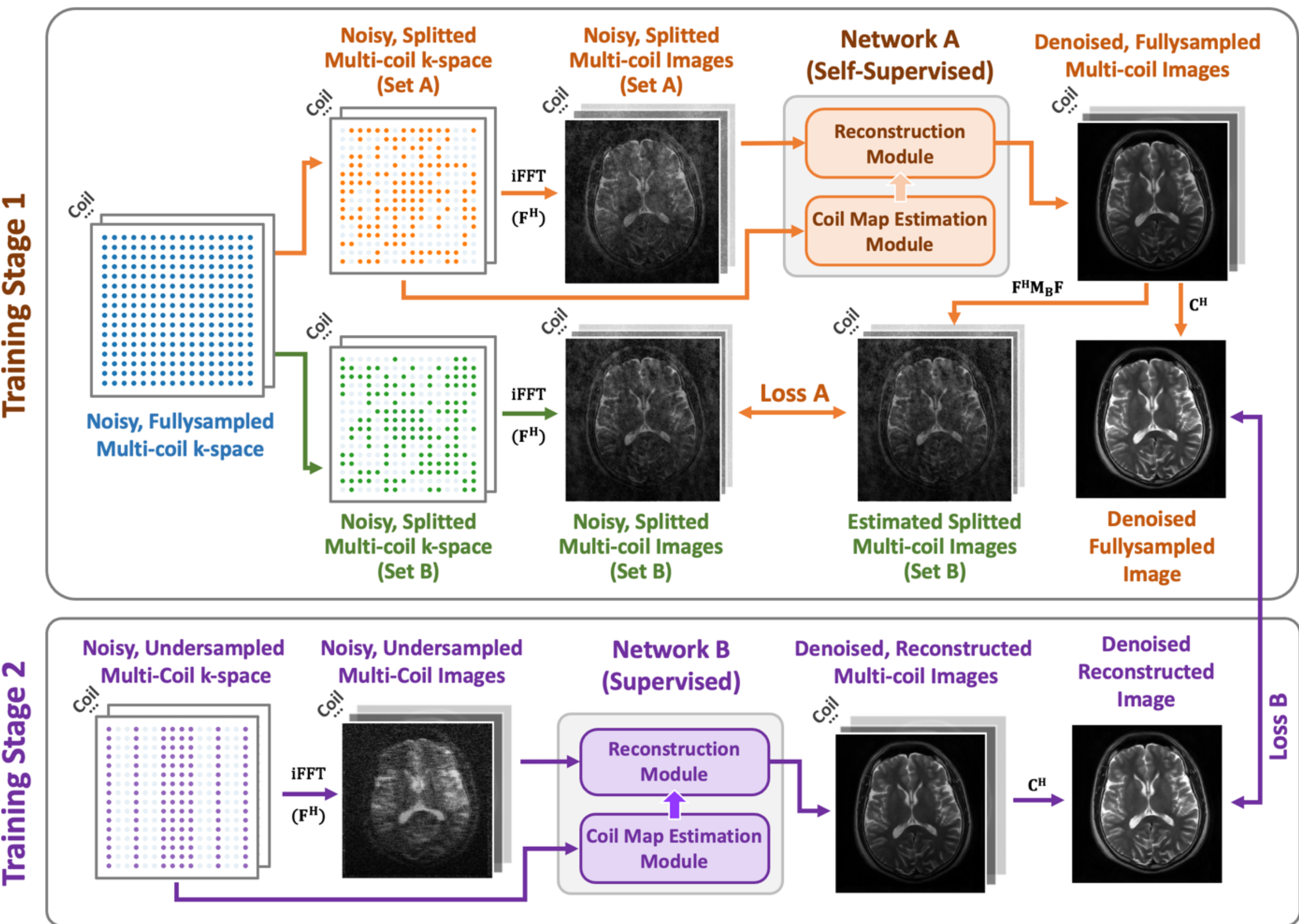

**Figure 2:** The hybrid learning pipeline for Cartesian sampling applications (e.g., Cartesian MRI of the Brain at 0.3T), following a similar training procedure as for non-Cartesian sampling applications.

The overall hybrid learning pipeline is illustrated in **Figures 1&2** for non-Cartesian (spiral) and Cartesian sampling applications, respectively. In the first training stage, a neural network $f_\theta$ (parameterized by $\theta$), referred to hereafter as Network A, is trained using a self-supervised learning scheme (Equation 5) on the fully sampled low-SNR data in $D^{(1)}$ to generate intermediate high-quality pseudo-reference images. After training, inference is performed on all k-space datasets in $D^{(1)}$ to generate associated pseudo-references for each sample:

$$\hat{x}_{\text{ref},j} = f_\theta\left(F^H W y_j^{(1)}\right), j = 1, \dots, N. \tag{6}$$

In the second training stage, another neural network $g_\phi$ (parameterized by $\phi$), referred to hereafter as Network B, is trained on $D^{(2)}$ using supervised learning (Equation 4), with the pseudo-references $\hat{x}_{\text{ref}}$ from the first stage serving as the training targets.

After training, Network B alone can be used to reconstruct high-quality images from new undersampled, low-SNR k-space data $y$ (unseen during training) as:

$$\hat{x} = g_{\phi}(F^{H}Wy) \tag{7}$$

In summary, the rationale for the improved performance of hybrid learning is as follows: the self-supervised training of Network A in the first stage can yield high-quality pseudo-references $\hat{x}_{\text{ref}}$ from low-SNR references. When pseudo-references are sufficiently close to the true references $x_{\text{ref}}$, the supervised training of Network B in the second stage becomes effectively equivalent to supervision with true references, which provides stronger guidance for effective joint reconstruction and denoising compared with purely supervised using noisy reference or purely self-supervised approaches when only low-SNR training references are available.

**Evaluation**

The performance of hybrid learning was evaluated in three experiments involving both non-Cartesian and Cartesian noisy datasets. Experiment 1 starts with a simulation study on breath-hold lung MRI at 3T acquired with a stack-of-spirals ultrashort echo time (spiral-UTE) sequence (39-44). Synthetic noise at varying levels was added to simulate noisy data and assess the performance of hybrid learning. While lung MRI at 3T can be suboptimal due to pronounced field inhomogeneity, it provides high-quality ground truth images, making it ideal for validating the reconstruction accuracy of our method. Experiment 2 then evaluated hybrid learning on real low-field breath-hold lung MRI acquired at 0.55T using the same spiral-UTE sequence, where fully sampled, high-SNR references cannot be obtained. This represents one of the main clinical targets of this study to improve lung MRI at 0.55T by leveraging the inherently improved field homogeneity at reduced field strengths. In order to test the generalizability of hybrid learning to Cartesian sampling, Experiment 3 tested its performance using real low-field brain MRI datasets acquired at 0.3T. In all experiments, hybrid learning was compared against two state-of-the-art baselines: (1) standard supervised learning using noisy references, and (2) self-supervised learning applied directly to undersampled noisy data.

***Experiment 1: Simulation Study on Breath-Hold Spiral MRI of the Lung at 3T with Synthetic Noise***

In this experiment, a total of 30 lung MRI datasets were acquired on a 3T MRI scanner (MAGNETOM Prisma, Siemens Healthineers, Erlangen, Germany) from 30 subjects (14 males, 16 females; mean age = 51 ± 16 years) for the simulation study. All participants provided written informed consent. For each subject, data was acquired during a single breath-hold using a spiral-UTE sequence with a uniform spiral sampling trajectory (39-41). MRI scans were performed in the coronal orientation with the following parameters: FOV = 480 × 480 $mm^2$, matrix size = 224 × 224, in-plane spatial resolution = 2.1 × 2.1 $mm^2$, TR/TE = 2.65 / 0.05 ms, number of slices = 48, flip angle = 5°, and the total scan time = 18 seconds. Each dataset comprised 140 spiral interleaves, with a readout duration of 1.16 ms per interleaf. This acquisition scheme produced images free of noticeable undersampling artifacts and is therefore treated as fully sampled (R = 1). After reconstruction, all images were interpolated to 96 slices with an isotropic voxel size of 2.1 $mm^3$.

Since the datasets were acquired at 3T, fully sampled, high-SNR references ($x_{\mathrm{ref}}$) were available. To simulate low-field conditions, zero-mean complex Gaussian noise was added to the original k-space measurements of all datasets at a predefined noise level ($\sigma^2$). 20 datasets were used for training and validation, and the remaining 10 datasets were used for evaluation. Three reconstruction methods, including supervised learning with noisy references, self-supervised learning, and hybrid learning, were applied separately to the simulated noisy datasets at two acceleration rates (R): R=2 (70 spiral interleaves, scan time = 9s) and R=3 (47 interleaves, scan time = 6s). Acceleration was achieved by regularly discarding spiral interleaves to yield uniform undersampling. The experiments were repeated for three noise levels: low ($\sigma^2 = 0.02$), medium ($\sigma^2 = 0.04$), and high ($\sigma^2 = 0.08$).

Reconstruction performance of all methods was assessed using the Structural Similarity Index Measure (SSIM) and Normalized Mean Square Error (NMSE), with the fully sampled, high-SNR images ($x_{\mathrm{ref}}$) serving as the ground truth. In addition, the fully sampled, denoised images ($\hat{x}_{\mathrm{ref}}$, the output from the first stage of the hybrid learning, referred to as pseudo-reference hereafter) were also used as a surrogate ground truth to evaluate whether such images could be used for quantitative validation in subsequent real low-field imaging experiments, where fully sampled, high-SNR references are not available.

***Experiment 2: Evaluation on Breath-Hold Spiral MRI of the Lung at 0.55T***

In this experiment, hybrid learning was evaluated on real-world 0.55T breath-hold spiral lung MRI, where fully sampled high-SNR references are unavailable. A total of 56 datasets were acquired from 56 subjects (24 males, 32 females; mean age 58 ± 13 years) on a ramped-down 0.55T MRI scanner (MAGNETOM Aera, Siemens Healthineers, Erlangen, Germany) using the same 3D spiral-UTE sequence as in Experiment 1. All participants provided written informed consent. MRI scans were performed in the coronal orientation with the following parameters: FOV = 500 × 500 $mm^2$, matrix size = 256 × 256, in-plane spatial resolution = 1.95 × 1.95 $mm^2$, TR/TE = 3.69 / 0.03 ms, number of slices = 64, flip angle = $5^o$, and the total scan time = 19 seconds. Each dataset contained 80 spiral interleaves, with a readout duration of 2.2 ms per interleaf. Similar to Experiment 1, this acquisition scheme produced images free of noticeable undersampling artifacts and is therefore treated as fully sampled (R = 1). After reconstruction, all images were interpolated to 128 slices with an isotropic voxel size of 1.95 $mm^3$.

41 datasets were used for training and validation, while the remaining 15 datasets were used for evaluation. The three reconstruction methods as in Experiment 1 were evaluated at R=2 (40 spiral interleaves, scan time = 9.5s) and R= 3 (27 interleaves, scan time = 6.3s). Because fully sampled high-SNR references are not available at 0.55T, the evaluation used fully sampled, denoised pseudo-reference images generated in the first stage of hybrid learning as a surrogate ground truth for computing SSIM and NMSE.

***Experiment 3: Evaluation on Cartesian MRI of the Brain at 0.3T***

In this experiment, we used brain datasets from the M4Raw database (45) to evaluate the generalizability of hybrid learning to Cartesian sampling. M4Raw contains real-world Cartesian brain datasets acquired at 0.3T, with detailed imaging parameters provided in the original reference (45).

A total of 128 fully sampled T2-weighted brain datasets from the M4Raw database were included in the study. 108 datasets were used for training and validation and the remaining 20 datasets were used for evaluation. Each dataset contains three separate repetitions. As described in the original paper (45), multi-repetition averaging can improve SNR by computing the magnitude average across the three repetitions. All three reconstruction methods from Experiments 1&2 were performed on the first repetition only of each dataset at R=2 and R=4 using a fixed 1D uniform undersampling pattern with 24 center of k-space lines.

Although multi-repetition averaging results in improved SNR, we observed that the averaged images still have residual noise, and this can affect quantitative evaluation. To obtain a cleaner and independent ground truth, we trained a separate self-supervised denoising model (identical to Stage 1 of hybrid learning) on the second and third repetition datasets. The denoised magnitude images from these two repetitions were then averaged to generate high-SNR reference images for computing SSIM and NMSE. Note that the second and third repetitions were only used to generate independent references for quantitative evaluation, and only the first repetition was used for training the hybrid learning model.

**Implementation**

All network weights were optimized using the adaptive gradient descent algorithm (ADAM) (46) with a learning rate of 1e-4 and a batch size of 1. Each unrolled network consisted of 12 cascaded blocks, resulting in 8.66 million trainable parameters per model. Training was performed for 200 epochs in PyTorch (version 2.0) on a server equipped with an NVIDIA Tesla A100 GPU. For supervised learning (including the second stage of hybrid learning), negative SSIM was used for the loss function, as suggested in (15). For self-supervised learning (including the first stage of hybrid learning), a normalized mixed L1-L2 loss was used following (21).

In self-supervised training (including the first stage of hybrid learning) for lung MRI reconstruction, a k-space splitting ratio between 0.3 and 0.99 was randomly assigned for each spiral interleaf, with the ratio updated at every backpropagation step. For brain MRI reconstruction with Cartesian sampling, the splitting ratio was set between 0.3 and 0.99 for the first stage of hybrid learning, while it was set to between 0.4 and 0.6 for self-supervised learning (excluding the first stage of hybrid learning), as recommended in (21) and confirmed in our experiments. In both cases, a small 4×4 central k-space region was shared between the two splitting sets to ensure consistent scaling across sets, as suggested in (47). Note that all splitting settings mentioned above were individually optimized to maximize performance for each method.

## Results

### Results of Experiment 1

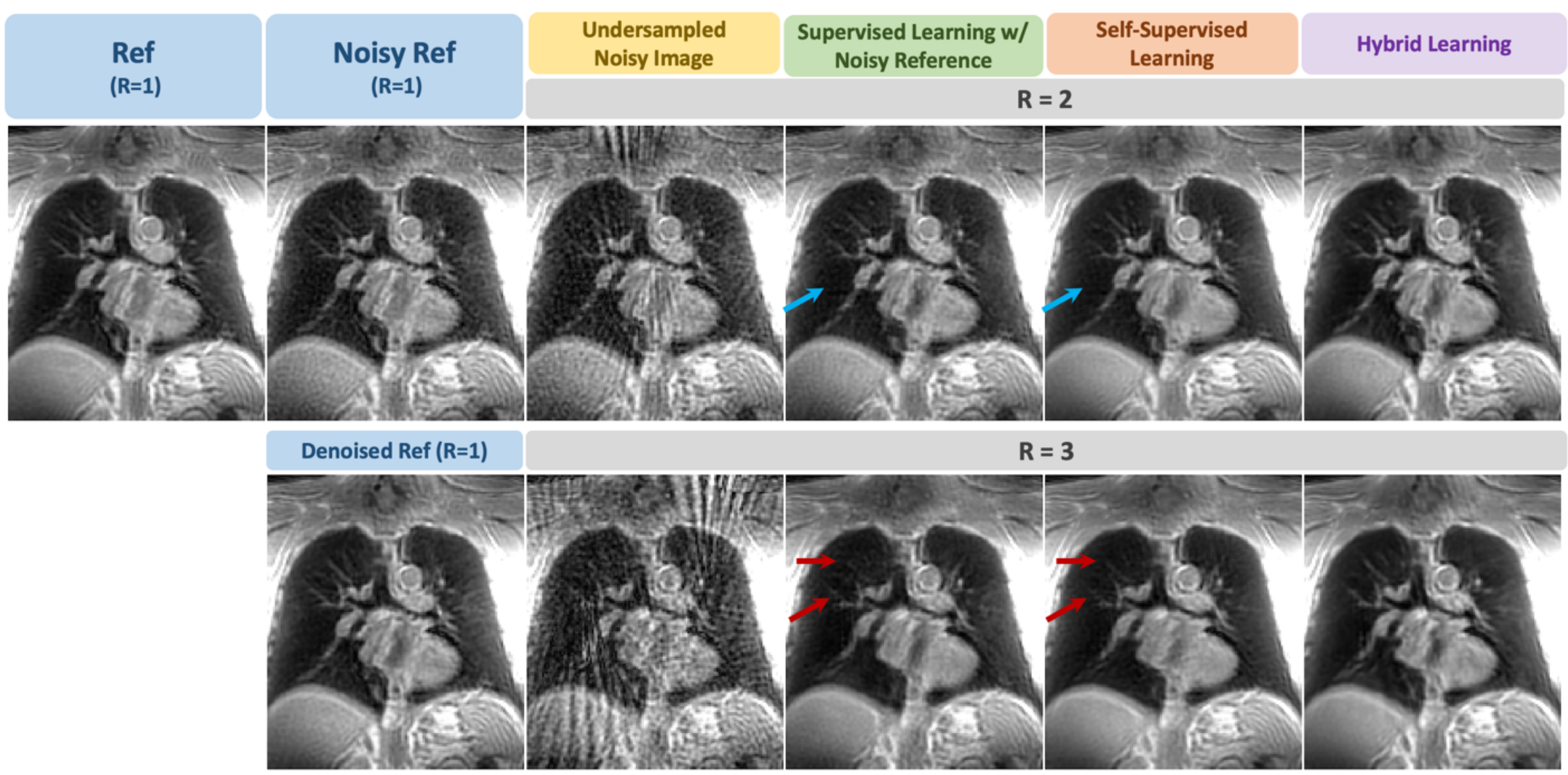

**Figure 3:** A representative case from the simulation study on 3T lung MRI at the low-noise level, comparing hybrid learning with supervised and self-supervised learning at R=2 and R=3. The original fully sampled image (Ref), the fully sampled noisy image with added noise (Noisy Ref), and the fully sampled denoised image (Denoised Ref) are included for comparison. Both supervised learning (trained with noisy references) and self-supervised learning exhibit residual noise (blue arrows) and structural loss (red arrows). In contrast, hybrid learning enables better reconstruction

quality and the results are closer to the ground truth. Note that the original fully sampled image (Ref) is only used for comparison in this experiment and was not used in any training process.

**Figure 3** shows a representative case from the simulation study on 3T lung MRI at the low-noise level, comparing hybrid learning with supervised and self-supervised learning at R=2 and R=3. The original fully sampled image (Ref), the fully sampled noisy image with added noise (Noisy Ref), and the fully sampled denoised image (Denoised Ref) are included for comparison. Both supervised learning (trained with noisy references) and self-supervised learning exhibit residual noise (blue arrows) and structural loss (red arrows). In contrast, hybrid learning enables better reconstruction quality and the results are closer to the ground truth. Note that the original fully sampled image (Ref) is only used for comparison in this experiment and was not used in any training process. **Supporting Information Figure S3 and Figure S4** provide results from the same case at medium- and high-noise levels, where hybrid learning also outperforms both supervised and self-supervised learning.

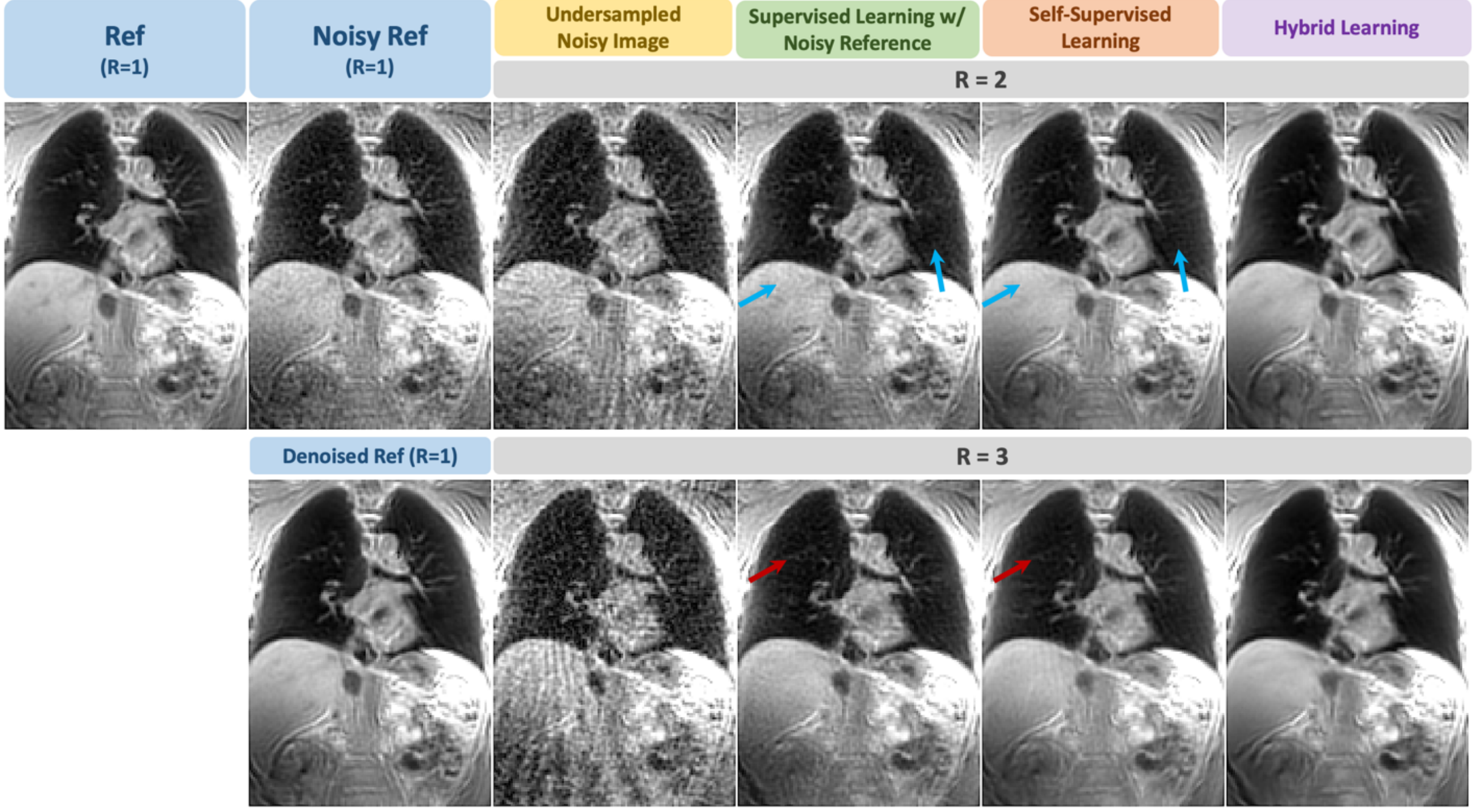

**Figure 4:** Another case from the simulation study on 3T lung MRI at the medium-noise level, comparing different reconstruction methods at R = 2 and R = 3. Both supervised and self-supervised

learning exhibit residual noise (blue arrows) and structural loss (red arrows), whereas hybrid learning produces improved image quality.

**Figure 4** presents another case at the medium-noise level, comparing different reconstruction methods at R = 2 and R = 3. Similar to the first case, both supervised and self-supervised learning exhibit residual noise (blue arrows) and structural loss (red arrows), whereas hybrid learning produces improved image quality. Consistent findings were observed for this case at low- and high-noise levels, as shown in **Supporting Information Figure S5 and Figure S6.**

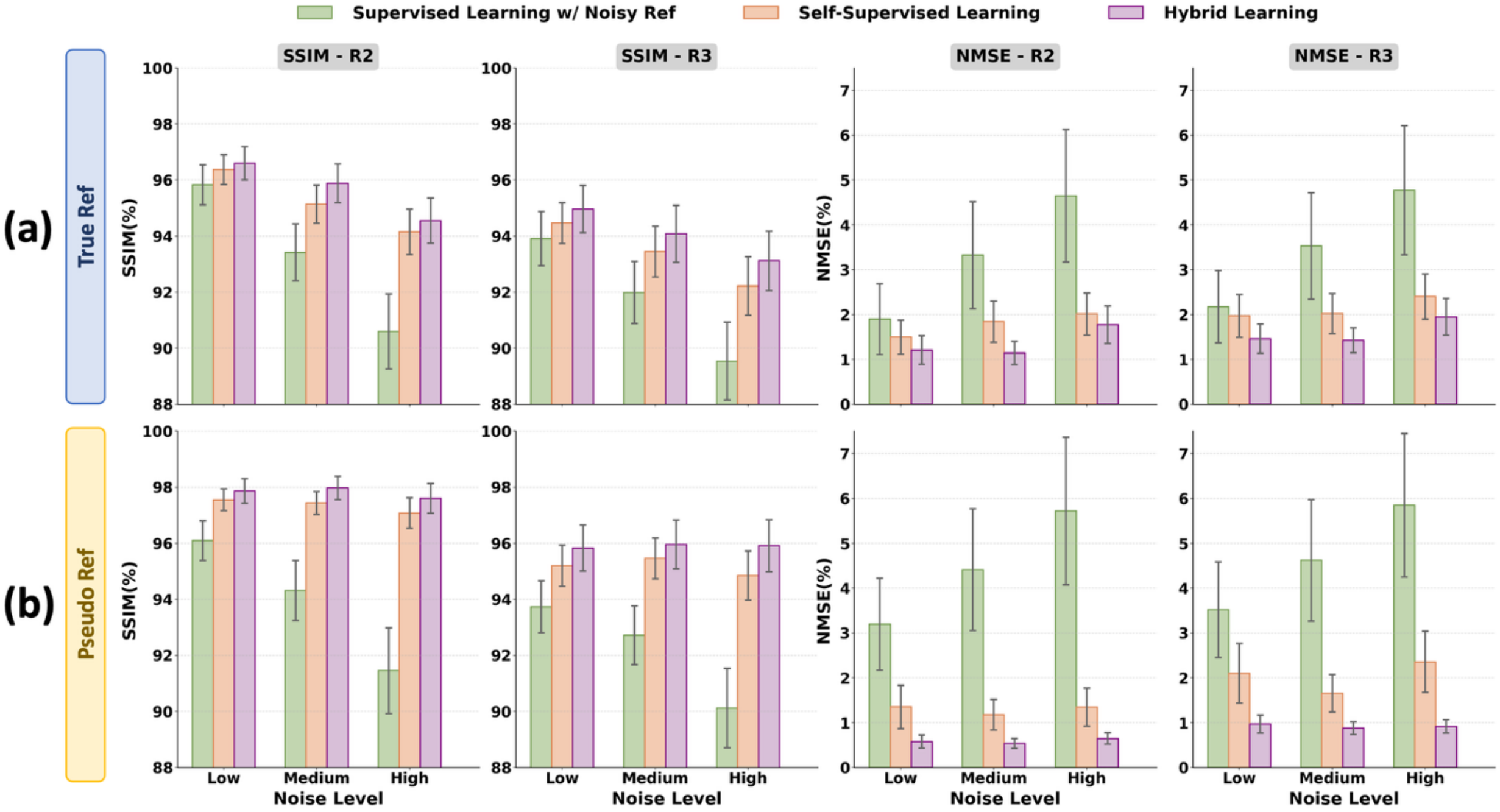

**Figure 5:** The quantitative comparison of different reconstruction methods, averaged across all testing cases (n = 10), using (a) the original fully sampled, high-SNR images for metric computation and (b) fully sampled, denoised pseudo-reference images for metric computation. The results show that hybrid learning achieved higher SSIM and lower NMSE than the other methods at both R = 2 and R = 3 across different noise levels. A similar trend was observed when the fully sampled, denoised pseudo-reference images were used for metric computation. The improvement of hybrid learning over the two reference methods was statistically significant ($P < 0.05$) for both SSIM and NMSE at both acceleration rates across different noise levels.

**Figure 5** presents the quantitative comparison of different reconstruction methods, averaged across all testing cases (n = 10). **Figure 5a** shows the results using the original fully sampled, high-SNR images as ground truth, where hybrid learning achieved higher SSIM and lower NMSE than the other methods at both R = 2 and R = 3 across different noise levels. A similar trend was observed when the fully sampled, denoised pseudo-reference images were used for metric computation, as shown in **Figure 5b**. The improvement of hybrid learning over the two reference methods was statistically significant ($P < 0.05$) for both SSIM and NMSE at both acceleration rates across different noise levels.

**Figures 4 and 5** show that the fully sampled, denoised pseudo-reference images are visually comparable to the original reference images prior to noise addition. To further validate this observation, additional quantitative comparisons were performed, as presented in **Supporting Information Figure S7**. These results suggest that, in real low-field imaging, fully sampled and denoised pseudo-reference images can serve as a reliable ground truth for both supervised training and quantitative evaluation without compromising accuracy.

**Results of Experiment 2**

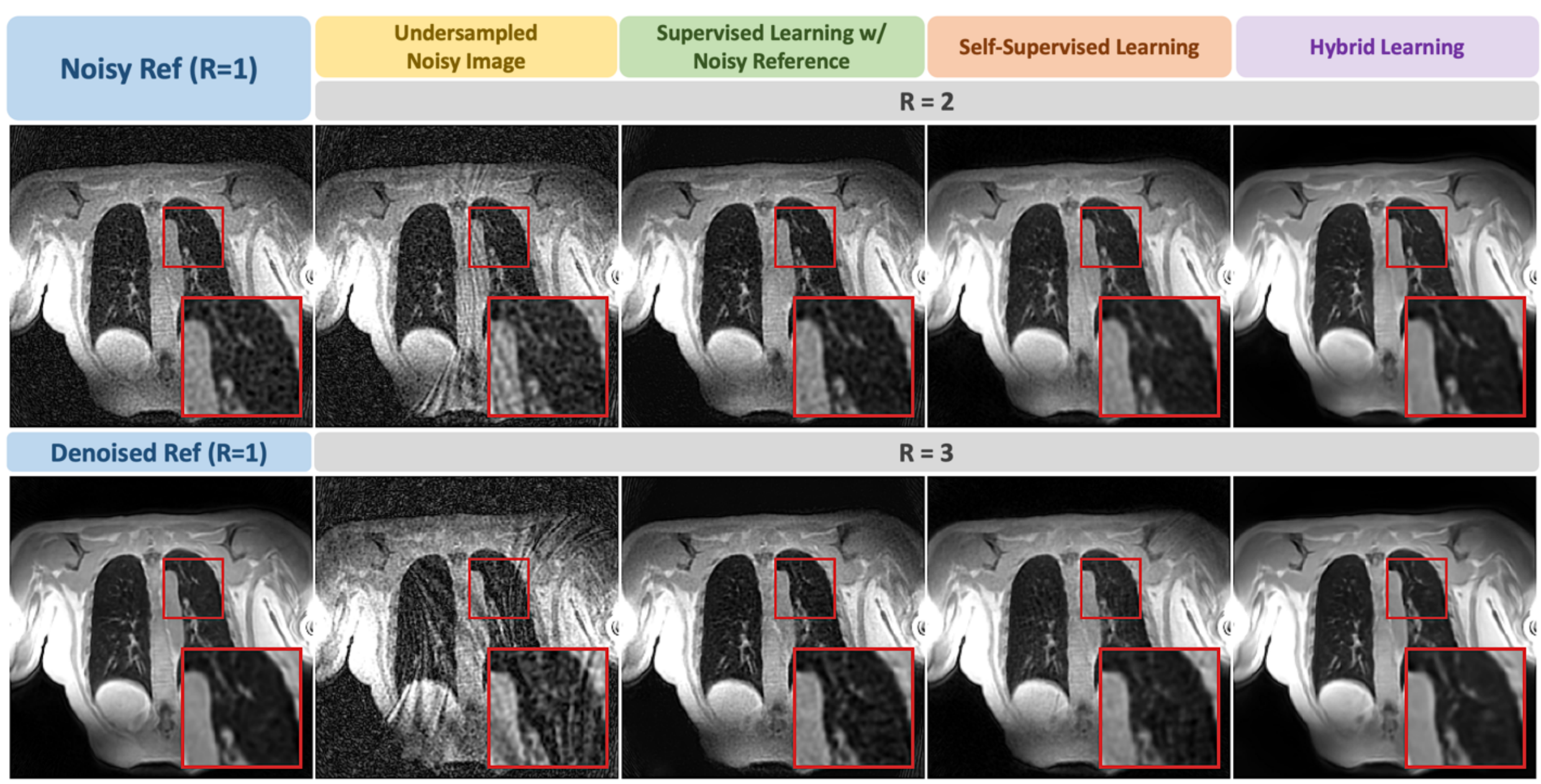

**Figure 6:** A representative case comparing hybrid learning with supervised and self-supervised learning at two acceleration rates for 0.55T lung MRI. At low field, the fully sampled image (Noisy Ref) exhibits noticeable noise, whereas the fully sampled denoised image (Denoised Ref) shows improved quality and visual SNR. At both acceleration rates, hybrid learning achieved superior image quality compared with the reference methods.

**Figure 6** shows a representative case comparing hybrid learning with supervised and self-supervised learning at two acceleration rates for 0.55T lung MRI. At low field, the fully sampled image (Noisy Ref) exhibits noticeable noise, whereas the fully sampled denoised image (Denoised Ref) shows improved quality and visual SNR. At both acceleration rates, hybrid learning achieved superior image quality compared with the reference methods.

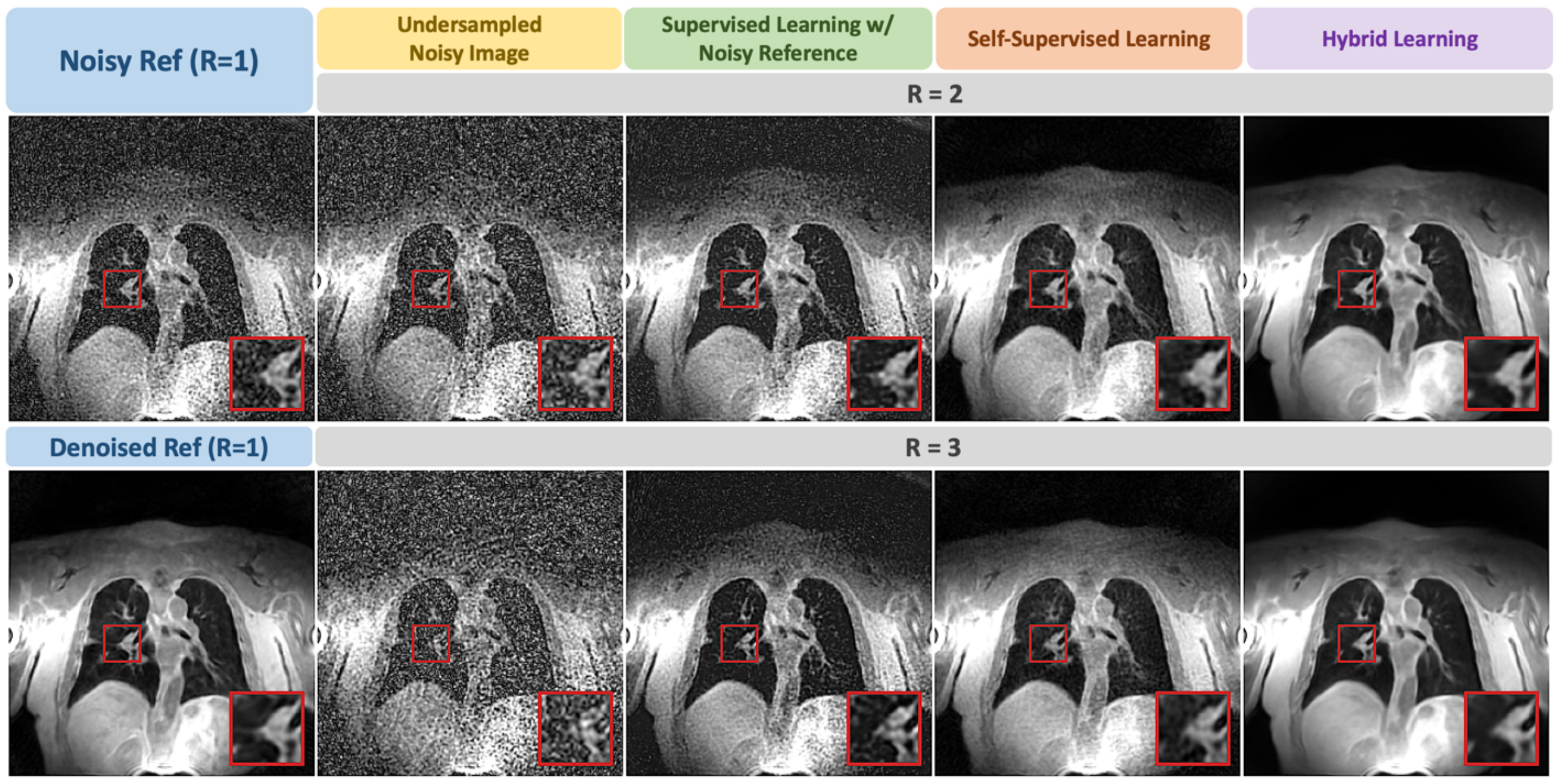

**Figure 7:** Another case with a higher noise level in the original image (Noisy Ref). In this case, hybrid learning maintained high reconstruction quality and effectively suppressed noise and undersampling artifacts, while both supervised and self-supervised learning produced compromised results, as highlighted in the zoomed-in region (red box).

**Figure 7** presents another case with a higher noise level in the original image (Noisy Ref). In this case, hybrid learning maintained high reconstruction quality and effectively suppressed noise and undersampling artifacts, while both supervised and self-supervised learning produced compromised results, as highlighted in the zoomed-in region (red box). **Supporting Information Figures S8 and S9** provide similar comparisons in additional two cases, where hybrid learning consistently yielded better image quality.

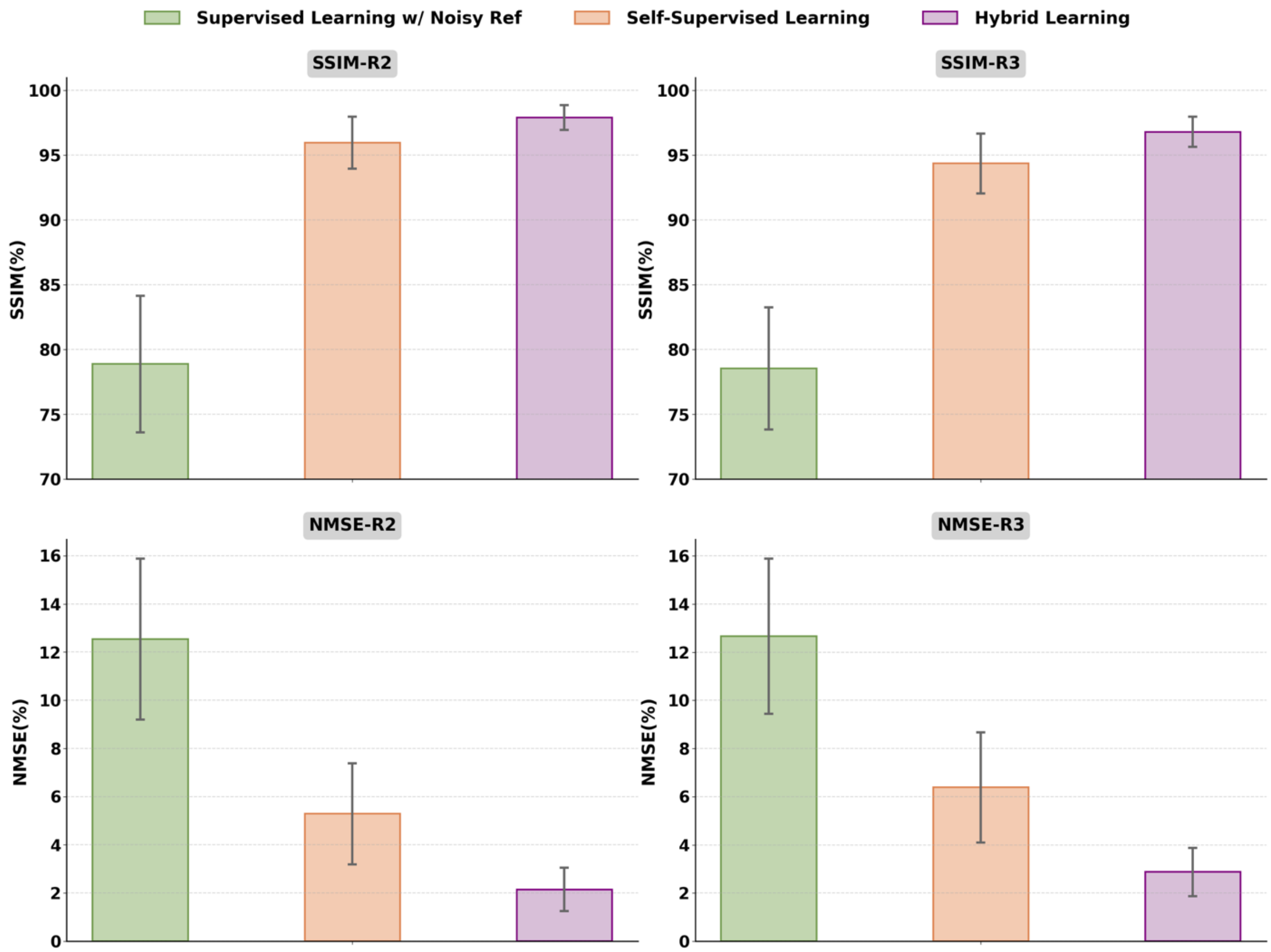

**Figure 8:** The quantitative comparison of different reconstruction methods across all testing cases (n=15) at R=2 and R=3, using the fully sampled denoised images as a reference after our validation in Experiment 1. Hybrid learning outperformed both supervised and self-supervised learning in terms of SSIM and NMSE, with the improvements reaching statistical significance (P<0.05) at both acceleration rates.

**Figure 8** shows the quantitative comparison of different reconstruction methods across all testing cases (n=15) at R=2 and R=3, using the fully sampled denoised images as a reference after our validation in Experiment 1. Hybrid learning outperformed both supervised and self-supervised learning in terms of SSIM and NMSE, with the improvements reaching statistical significance (P<0.05) at both acceleration rates.

## Results of Experiment 3

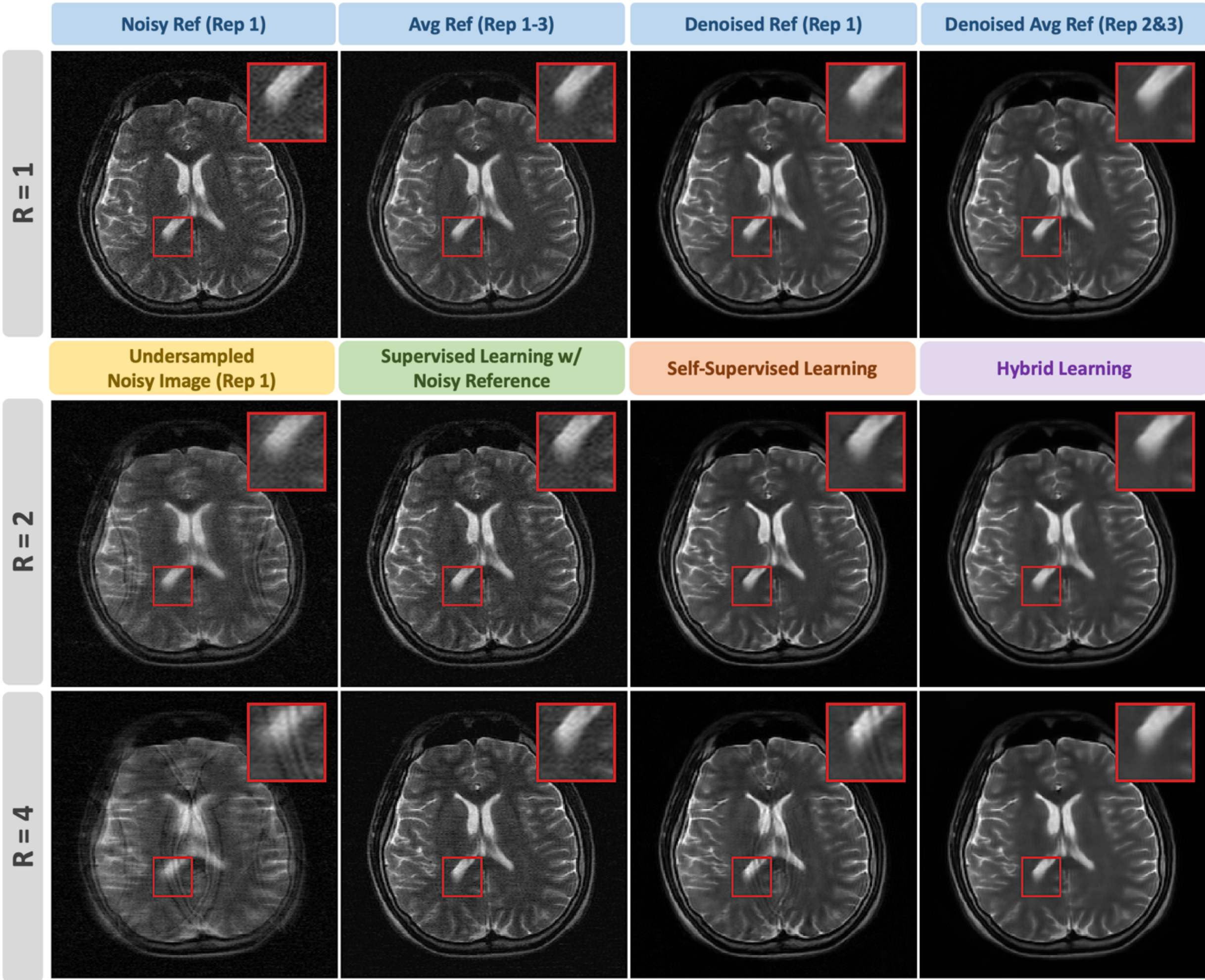

**Figure 9:** A representative case of 0.3T brain MRI with Cartesian sampling, comparing hybrid learning with supervised and self-supervised learning at acceleration rates R=2 and R=4. The top row shows fully sampled images under different conditions, including: (i) the original noisy image from the first repetition, (ii) the averaged image across all three repetitions, (iii) the denoised image from the first repetition, and (iv) the denoised averaged image from the second and third repetitions. At both R=2 and R=4, reconstructions from standard supervised and self-supervised learning on the first repetition fail to fully suppress residual noise. In contrast, hybrid learning provides clearer anatomical detail and more effective suppression of noise and artifacts, as highlighted in the zoomed-in regions (red boxes).

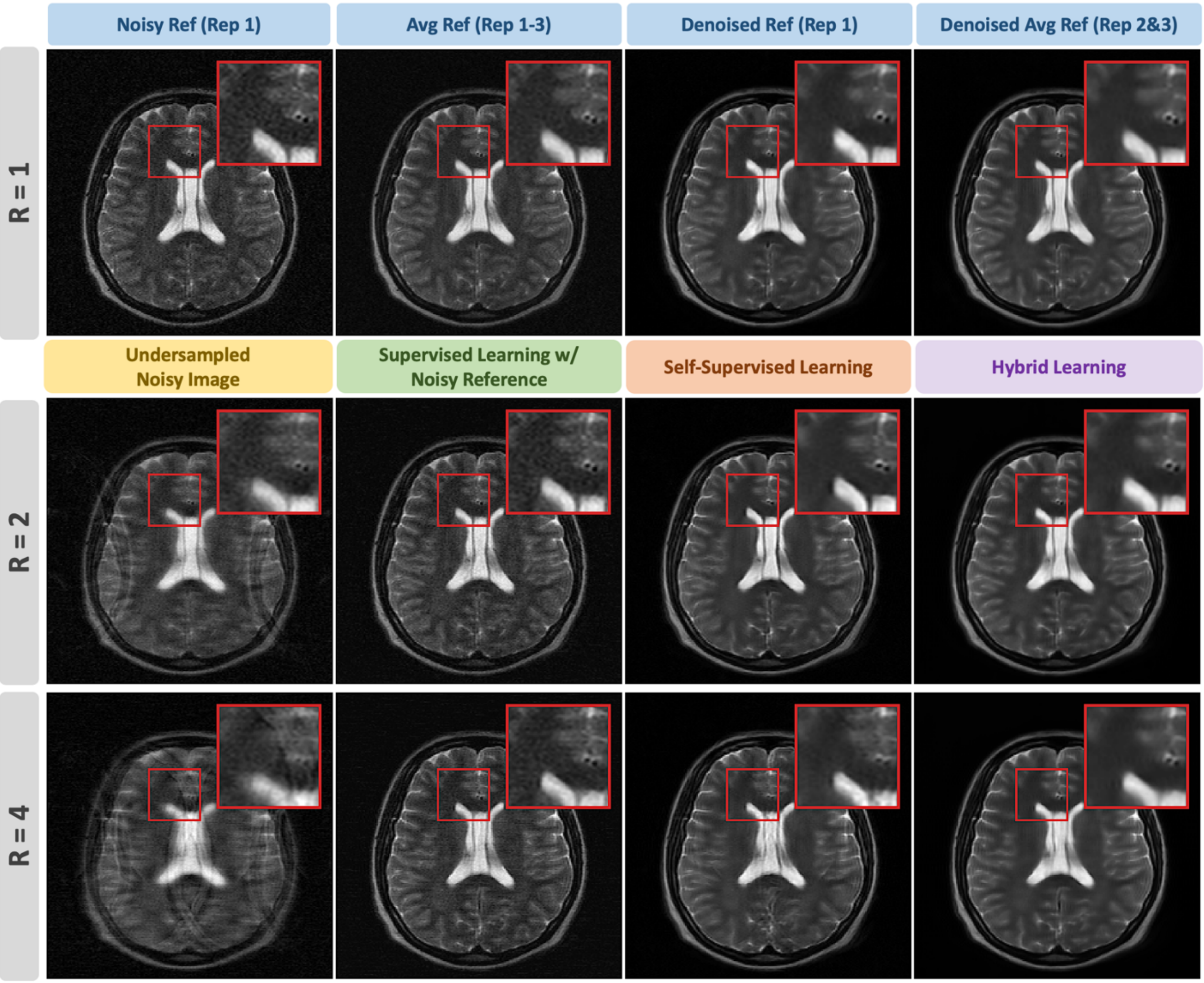

**Figure 10:** A similar comparison in another case from the 0.3T brain MRI with Cartesian sampling. Hybrid learning also achieves better visual image quality with reduced noise and residual artifacts.

**Figure 9** presents a representative case of 0.3T brain MRI with Cartesian sampling, comparing hybrid learning with supervised and self-supervised learning at acceleration rates R=2 and R=4. The top row shows fully sampled images under different conditions, including: (i) the original noisy image from the first repetition, (ii) the averaged image across all three repetitions, (iii) the denoised image from the first repetition, and (iv) the denoised averaged image from the second and third repetitions. At both R=2 and R=4, reconstructions from standard supervised and self-supervised learning on the first repetition fail to fully suppress residual noise. In contrast, hybrid learning provides clearer anatomical detail and more effective suppression of noise and artifacts, as highlighted in the zoomed-in regions (red boxes). **Figure 10** shows a similar comparison in another case, where hybrid learning also achieves better visual image quality with reduced noise and residual artifacts.

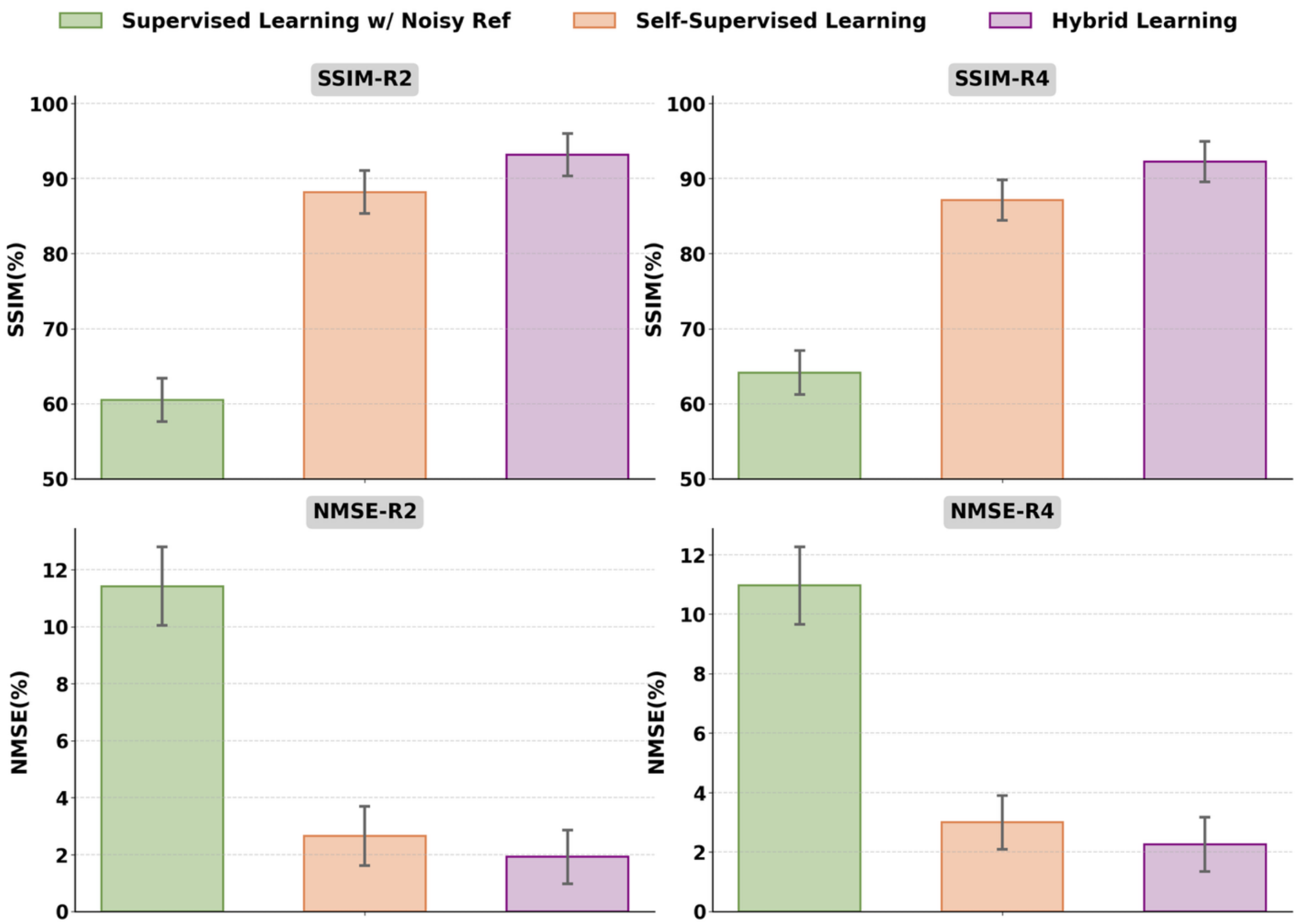

**Figure 11:** The quantitative comparison of different reconstruction methods across all testing cases (n = 20) at acceleration rates R = 2 and R = 4. Metrics were computed using independent reference images (Denoised Avg Ref, Rep 2&3) as described in the Methods. Hybrid learning outperformed both supervised and self-supervised learning in terms of SSIM and NMSE at both acceleration rates, with improvements reaching statistical significance ($P < 0.05$) for both metrics.

**Figure 11** shows the quantitative comparison of different reconstruction methods across all testing cases (n = 20) at acceleration rates R = 2 and R = 4. Metrics were computed using independent reference images (Denoised Avg Ref, Rep 2&3) as described in the Methods. Hybrid learning outperformed both supervised and self-supervised learning in terms of SSIM and NMSE at both acceleration rates, with improvements reaching statistical significance ($P < 0.05$) for both metrics.

## Discussion

In this study, we proposed hybrid learning, a two-stage deep learning framework that effectively integrates self-supervised learning in the first stage with supervised learning in the second stage for joint MRI reconstruction and denoising. This training scheme addresses a common challenge in low-field MRI reconstruction, where high-SNR reference images are often difficult to acquire for network training. While all practical MRI data contain noise and biases, standard high-field acquisitions typically provide sufficiently high SNR for conventional learning-based methods to perform well, whereas low-field imaging often suffers from substantially lower SNR, which highlights the need for more robust strategies.

The proposed hybrid learning technique first generates higher-quality pseudo-references from fully sampled noisy data through self-supervised learning, and then uses these pseudo-references for supervised training in the second step. Our results demonstrate that hybrid learning consistently outperforms both standard supervised learning and self-supervised learning alone when the baseline SNR is low. This is expected, as noisy training references limit the performance of conventional supervised approaches, while standard self-supervised methods tend to degrade when the baseline SNR is reduced.

The performance of hybrid learning was evaluated in three complementary experiments. A common challenge in assessing low-field MRI denoising and reconstruction techniques is the lack of reliable high-SNR ground truth. Experiment 1 addressed this issue through a simulation study on 3T spiral lung MRI, where fully sampled high-SNR images were available as ground truth. Synthetic noise was added to simulate low-field conditions to enable quantitative validation of our method across different noise levels. Although lung MRI at 3T suffers from pronounced field inhomogeneity and may be clinically suboptimal, it provided a practical way for us to test the accuracy of our approach and to evaluate the feasibility of using fully sampled, denoised images (results from the first stage of hybrid learning) as surrogate references when high-SNR ground truth is not available.

In Experiment 2, we evaluated hybrid learning on real-world 0.55T spiral lung MRI. Low-field imaging at 0.55T offers improved field homogeneity, making it a promising platform for lung imaging. However, the intrinsically lower SNR at reduced field strengths can substantially degrade image quality. Hybrid learning effectively suppressed both noise and undersampling artifacts, enabling shorter breath-hold durations while preserving image quality. Because of the lack of fully sampled, high-SNR ground truth, quantitative assessment in this experiment was performed using fully sampled, denoised images as pseudo-references, following the validation of this approach in Experiment 1.

In Experiment 3, we tested the generalizability of hybrid learning to Cartesian sampling using the public M4Raw brain MRI dataset acquired at 0.3T. The M4Raw datasets provide three independent repetitions. This allows us to evaluate our proposed method using the first repetition data while using the other two repetitions as independent references for quantitative evaluation. The results confirmed that hybrid learning also outperforms standard supervised and self-supervised learning approaches in different organs, field strengths, and sampling trajectories.

Hybrid learning is not limited to a specific self-supervised training strategy or network architecture. In this work, we implemented SSDU for self-supervised learning in the first stage, but other approaches, such as Equivariant Imaging (EI) (24, 31) or generative models (18), could be used, either independently or in combination with SSDU. Similarly, network architectures beyond the unrolled network implemented in this work could also be explored (18, 48), depending on the application and acquisition protocol.

Several prior studies have addressed MRI reconstruction under low-SNR conditions, which are closely related to our work and warrant discussion. Desai et al. proposed Noise2Recon (49), which improves SNR robustness for MRI reconstruction but still relies on access to high-SNR data and synthetic noise augmentation during training. Millard et al. proposed an extension of SSDU, termed Robust SSDU (37), to enhance self-supervised learning in low-SNR regimes. However, as a purely self-supervised method, its performance can degrade at high acceleration factors, particularly under fixed undersampling patterns. More recently Aali et al. proposed a pipeline for reconstruct undersampled data with low SNR (50), in which a self-supervised denoiser was first used to generate clean references for subsequent supervised reconstruction. While conceptually similar to our hybrid learning strategy, their approach employed Stein's Unbiased Risk Estimator (SURE) (51) as a pre-processing step for denoising. This requires prior knowledge of the noise level and is highly sensitive to noise estimation errors, which can lead to suboptimal performance (52). Moreover, their method was only validated on simulated data, leaving its applicability to real low-field MRI untested. In contrast, our approach incorporates denoising directly into the first step as a joint reconstruction–denoising process without the need for explicit noise-level estimation. Importantly, we demonstrate the efficacy of our method on real low-field datasets acquired with both Cartesian and non-Cartesian sampling.

While the current implementation of hybrid learning employs a sequential training scheme, it can accommodate alternative training strategies. One potential approach is pre-trained weight initialization, where the weights of the second-stage network could be initialized with the pre-trained weights from the first-stage network. Another alternative is joint training of both training stages, where instead of training the two stages sequentially, a parallel training strategy could be implemented. In this approach, the self-supervised loss from the first stage and the supervised loss from the second stage would be combined to jointly update either two separate networks or a single unified network.

Although this study focused on low-field MRI applications, the hybrid learning framework can be extended to other applications. For example, in dynamic MRI (53-55), where fully sampled references are hard to obtain due to continuous motion, hybrid learning could provide a robust solution for motion-resolved and motion-compensated reconstructions combining self-supervised learning and supervised training in two training stages.  In addition, hybrid learning can also be applied to accelerated quantitative MRI applications (56-62), where fully sampled references are also hard to obtain due to extended scan time. Another potential application is super-resolution MRI, where hybrid learning could help bridge the gap between low-resolution and high-resolution image domains to improve spatial resolution while preserving fine structural details. These extensions will be explored in future work.

Despite its advantages, a major challenge of hybrid learning is the potential motion inconsistency between the output of the first training stage (e.g., fully sampled denoised images) and the input of the second stage (accelerated noisy images). If these images are misaligned due to motion, the reconstruction may suffer from blurring. This was not an issue in our study, which focused on breath-hold lung imaging and brain imaging, but it is likely to be problematic in free-breathing applications. One potential solution is to modify the second stage to incorporate self-supervised learning, using the first-stage reconstructions as guidance rather than direct supervision. This approach could enable the second-stage network to adapt to motion variations while still leveraging the structural information provided in the first phase. Future studies will be needed to evaluate such strategies and improve the robustness of hybrid learning in motion-sensitive MRI applications.

## Conclusion

This work proposes a hybrid learning framework for joint MRI denoising and reconstruction for applications in low-field MRI, where acquiring high-SNR reference datasets for network training can be challenging. Compared to standard supervised and self-supervised learning, hybrid learning demonstrates superior image reconstruction quality with more accurate and robust results. The framework holds strong potential for a wide range of low-field MRI applications.

## Acknowledgements

This work was supported in part by the NIH (R01EB030549, EB031083, R21EB032917, and P41EB017183) and was performed under the rubric of the Center for Advanced Imaging Innovation and Research (CAI$^2$R), an NIBIB National Center for Biomedical Imaging and Bioengineering. The authors would like to thank Mary Bruno and Mahesh Keerthivasan for their support with spiral lung MRI scans.

## Supporting Information Figures

Description: Additional figures for the unrolled network structure and additional results.

### Figure S1

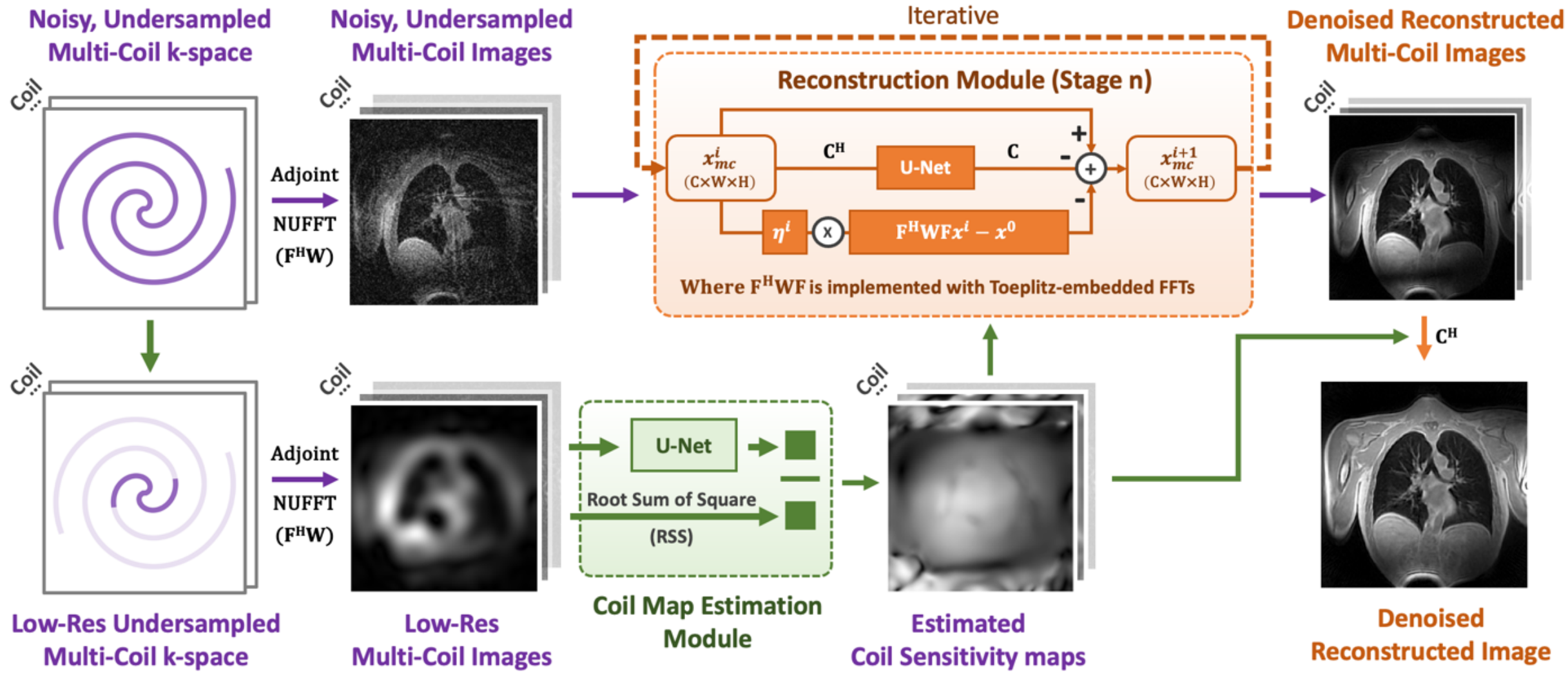

The architecture of the unrolled network for non-Cartesian (spiral) MRI reconstructions (e.g., breath-hold Spiral MRI of the Lung at 0.55T). This unrolled network consists of a reconstruction module and a coil sensitivity estimation module. The reconstruction module employs multiple small U-Nets to model iterative gradient descent updates, where a CNN estimates the gradient of the regularization function. Meanwhile, the coil sensitivity estimation module uses another small U-Net to estimate coil sensitivity maps from the center of the k-space data, which are then incorporated into the reconstruction process.

**Figure S2**

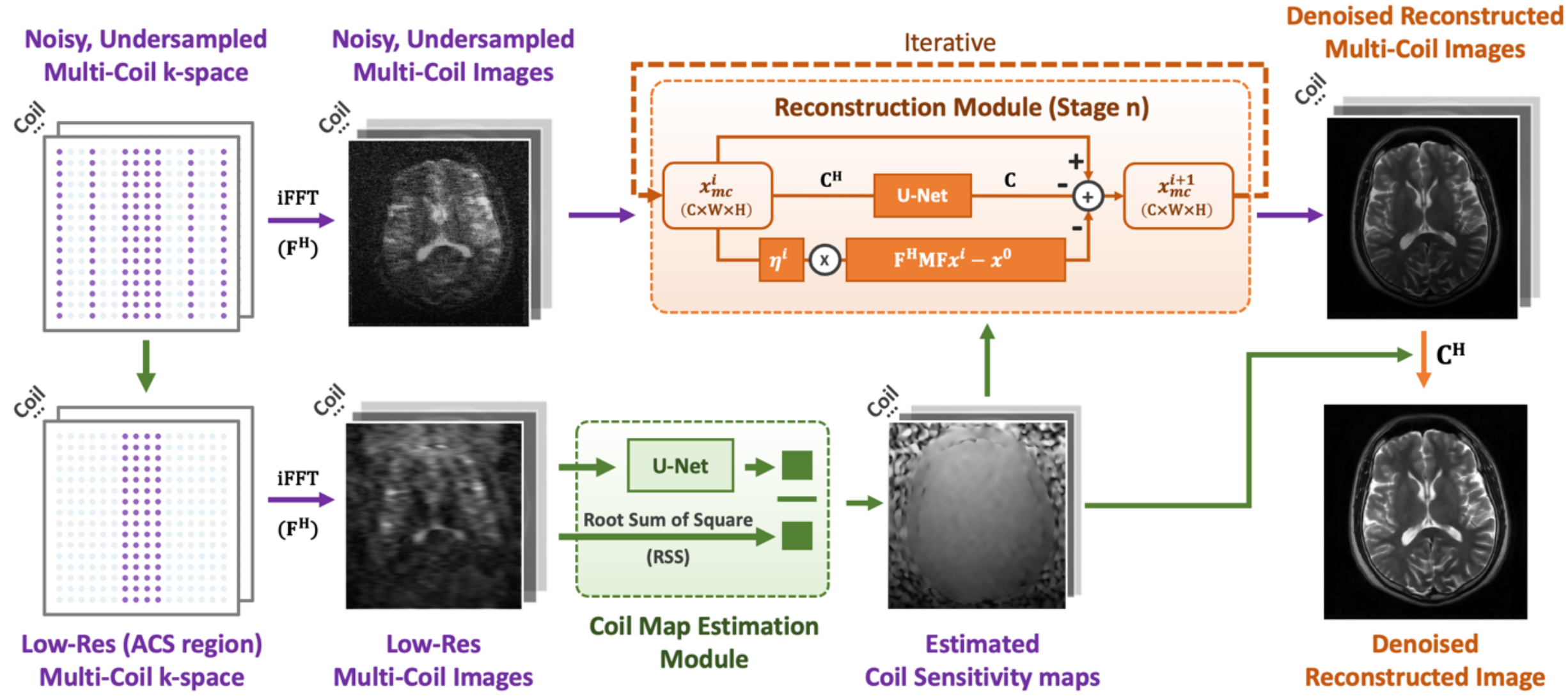

The architecture of the unrolled network for Cartesian MRI reconstructions (e.g., Cartesian MRI of the Brain at 0.3T). This unrolled network consists of a reconstruction module and a coil sensitivity estimation module. The reconstruction module employs multiple small U-Nets to model iterative gradient descent updates, where a CNN estimates the gradient of the regularization function. Meanwhile, the coil sensitivity estimation module uses another small U-Net to estimate coil sensitivity maps from the center of the k-space, which are then incorporated into the reconstruction process.

**Figure S3**

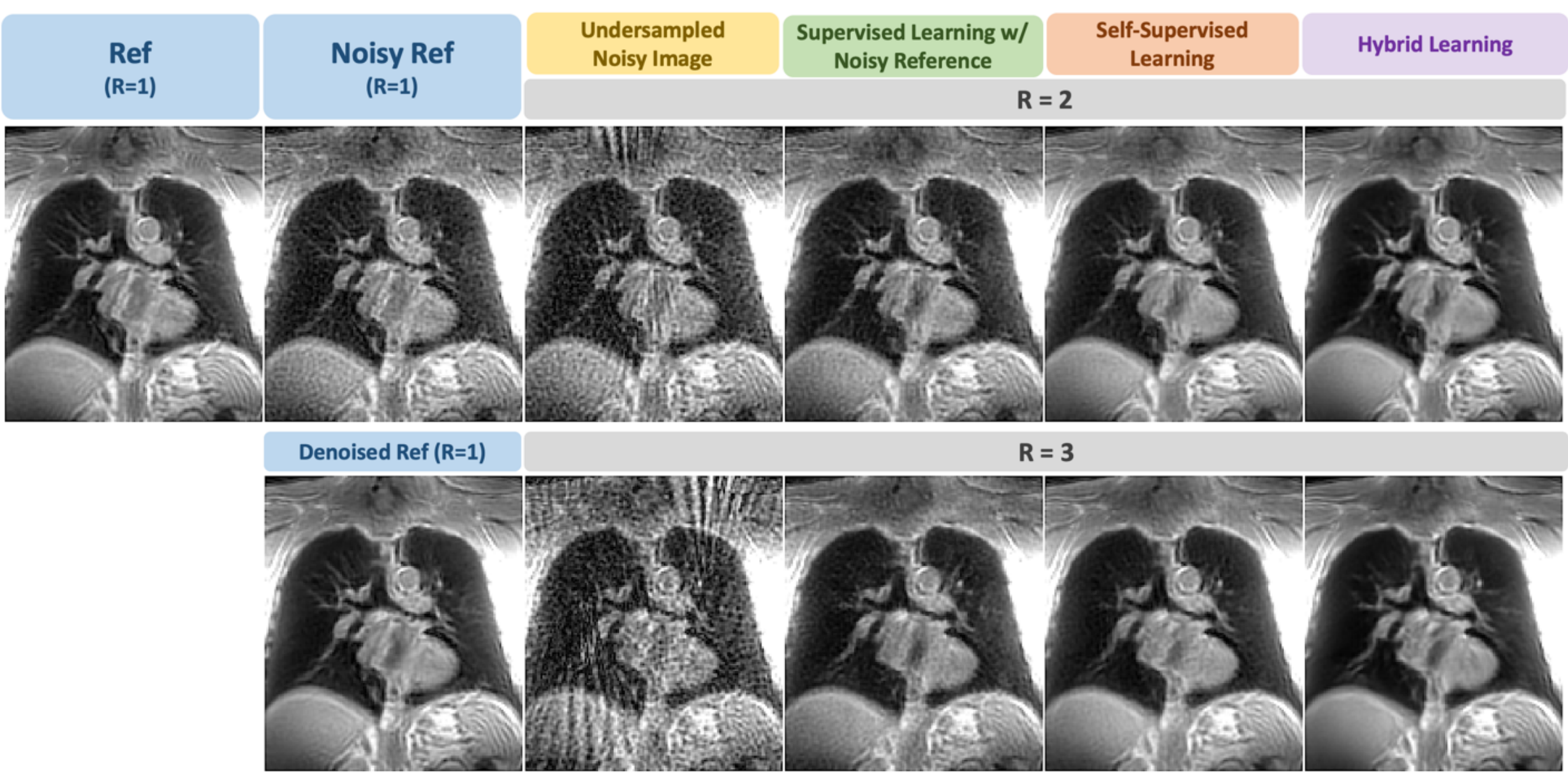

Results from the same case shown in Figure 3 at medium-noise levels, where hybrid learning outperforms both supervised and self-supervised learning.

**Figure S4**

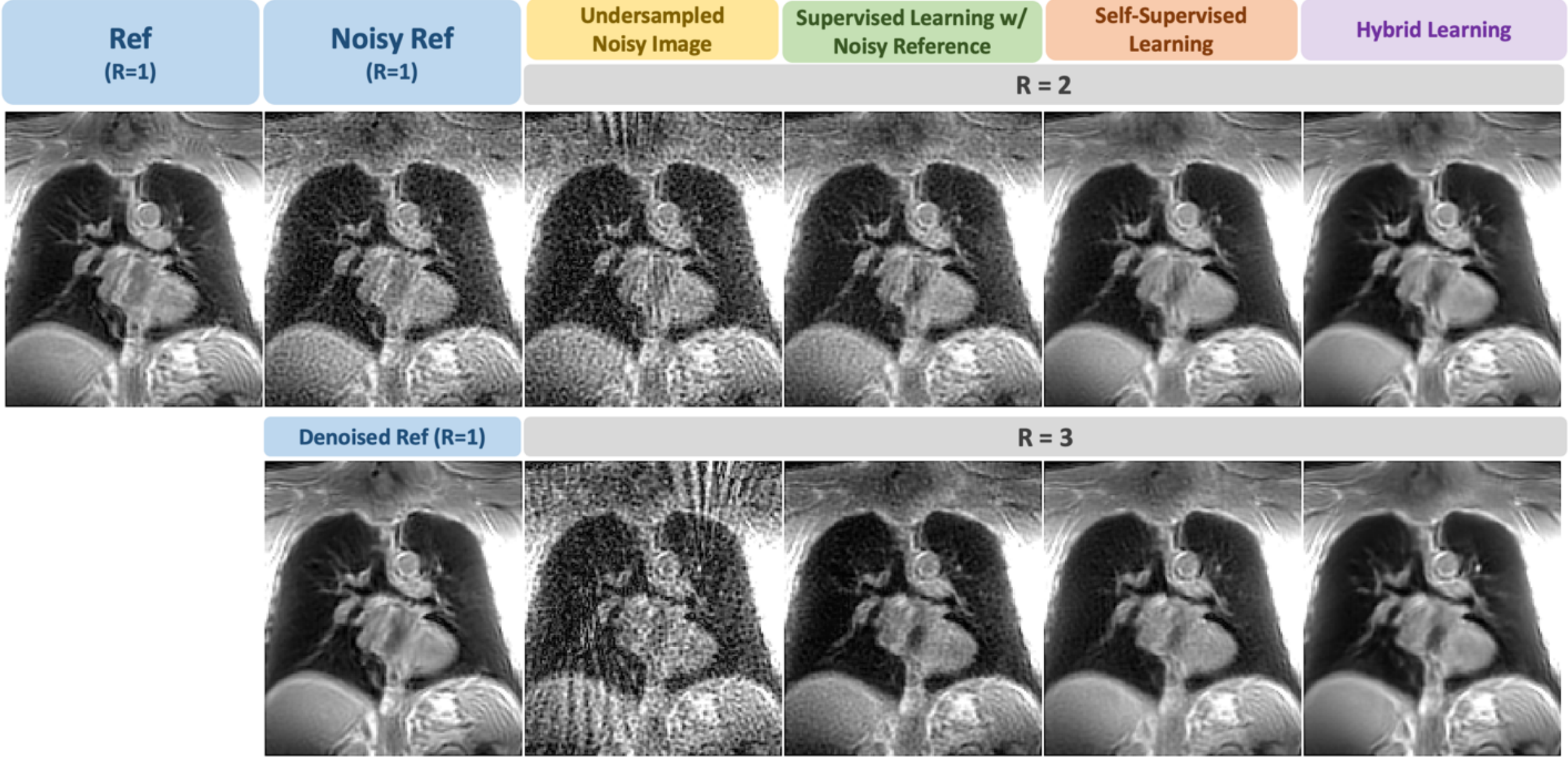

Results from the same case shown in Figure 3 at high-noise levels, where hybrid learning outperforms both supervised and self-supervised learning.

**Figure S5**

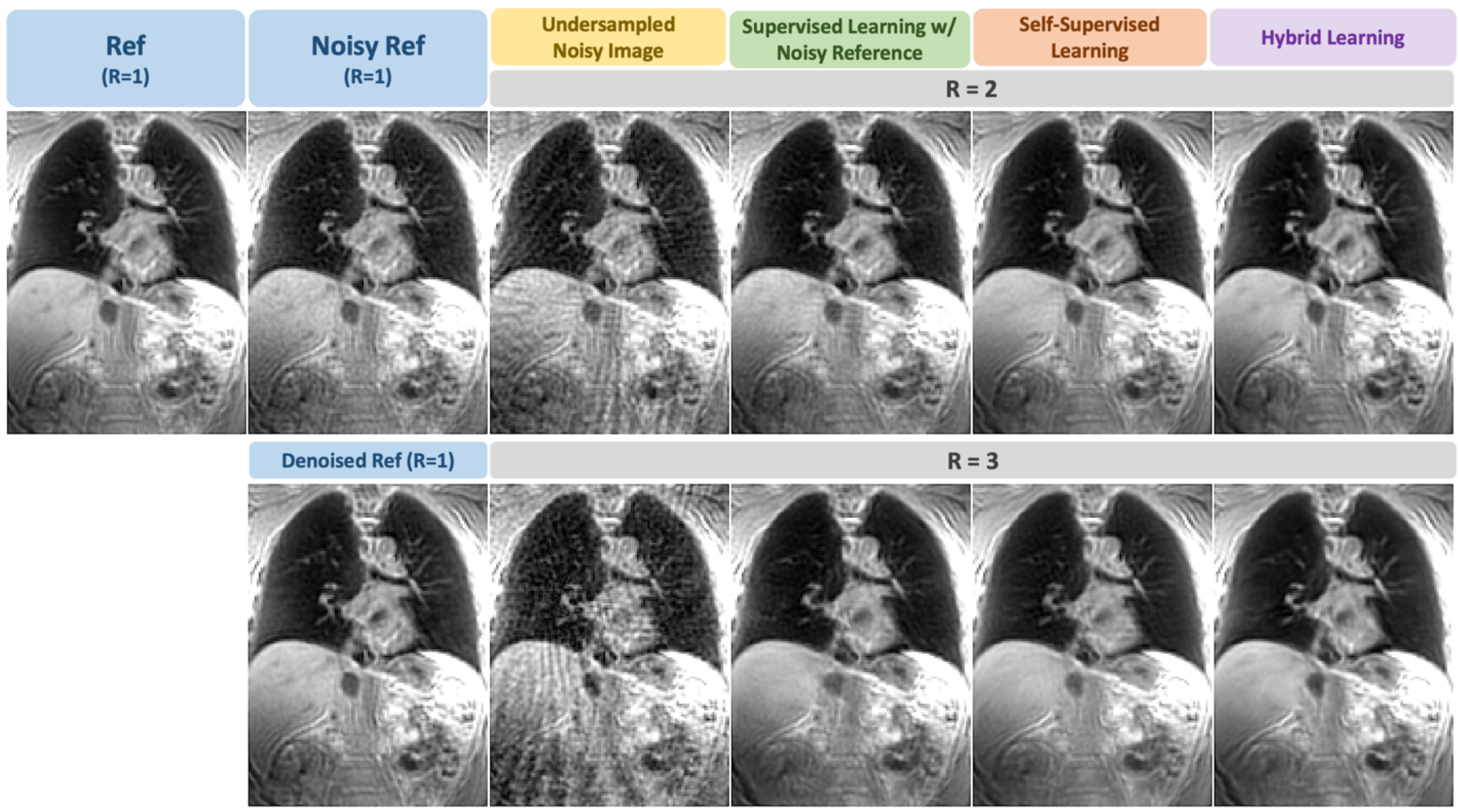

Results from the same case shown in Figure 4 at low-noise levels, where hybrid learning outperforms both supervised and self-supervised learning.

**Figure S6**

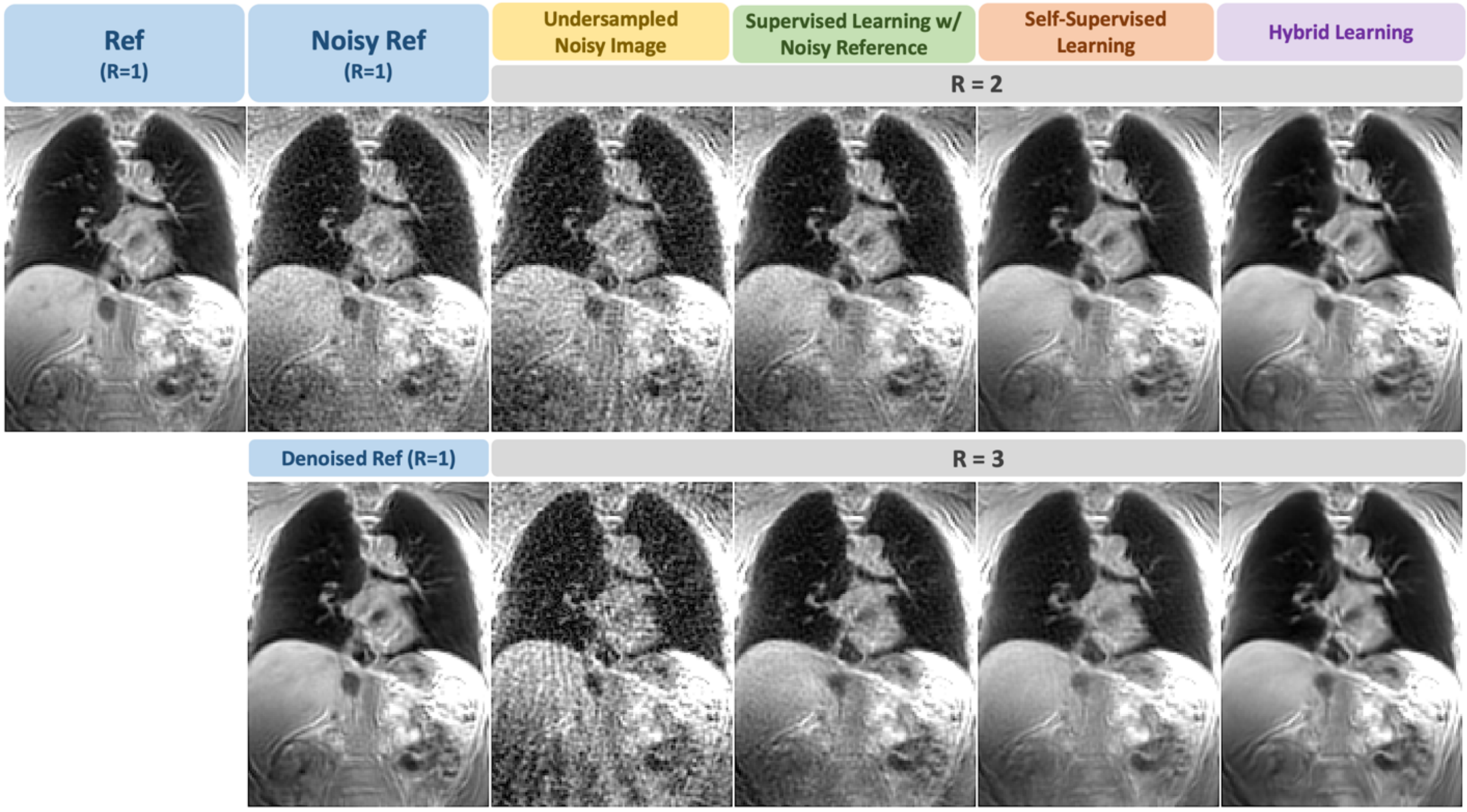

Results from the same case shown in Figure 4 at high-noise levels, where hybrid learning outperforms both supervised and self-supervised learning.

**Figure S7**

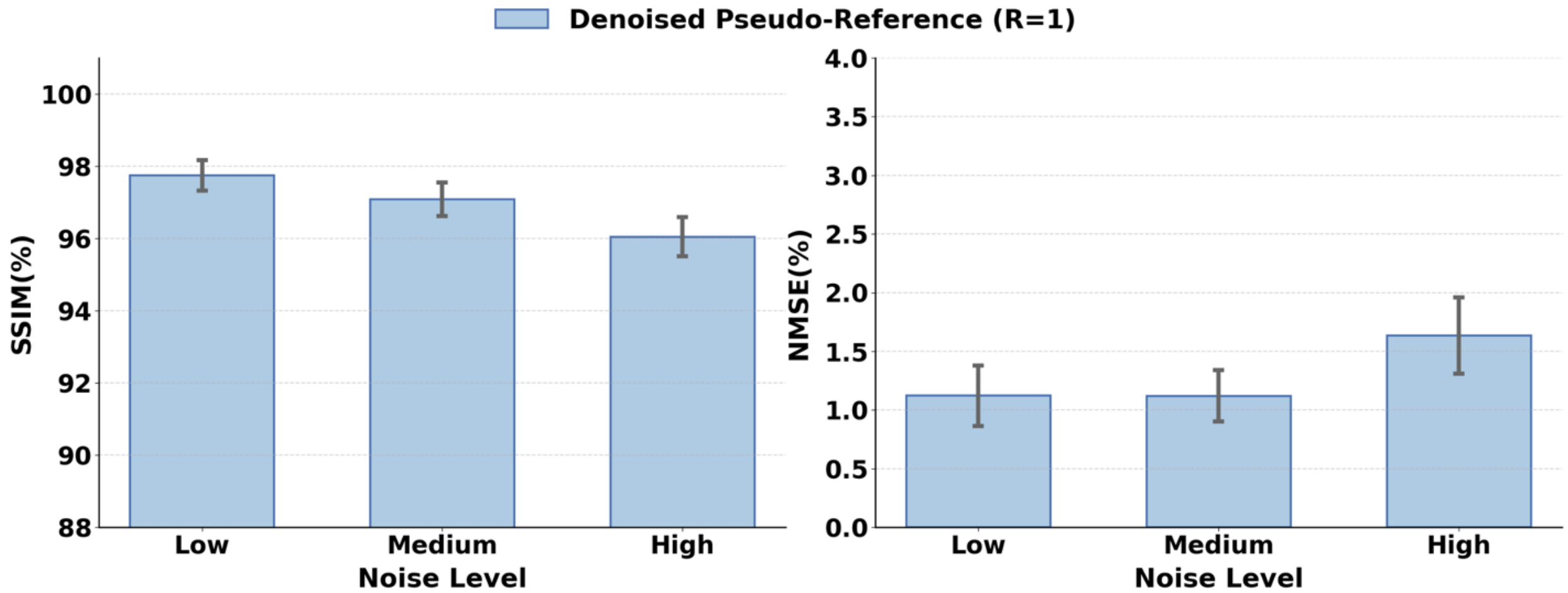

The quantitative measurement of denoised pseudo-references generated from the first stage of the hybrid learning across all testing cases (n = 10) from the simulation study on 3T lung MRI at different noise levels. We found that the fully sampled, denoised pseudo-reference images closely matched the original images prior to noise addition, both visually and in quantitative assessment. These results suggest that, in real low-field imaging, fully sampled, denoised pseudo-reference images can serve as a reliable ground truth for the supervised training and quantitative evaluation without compromising accuracy.

**Figure S8-9**

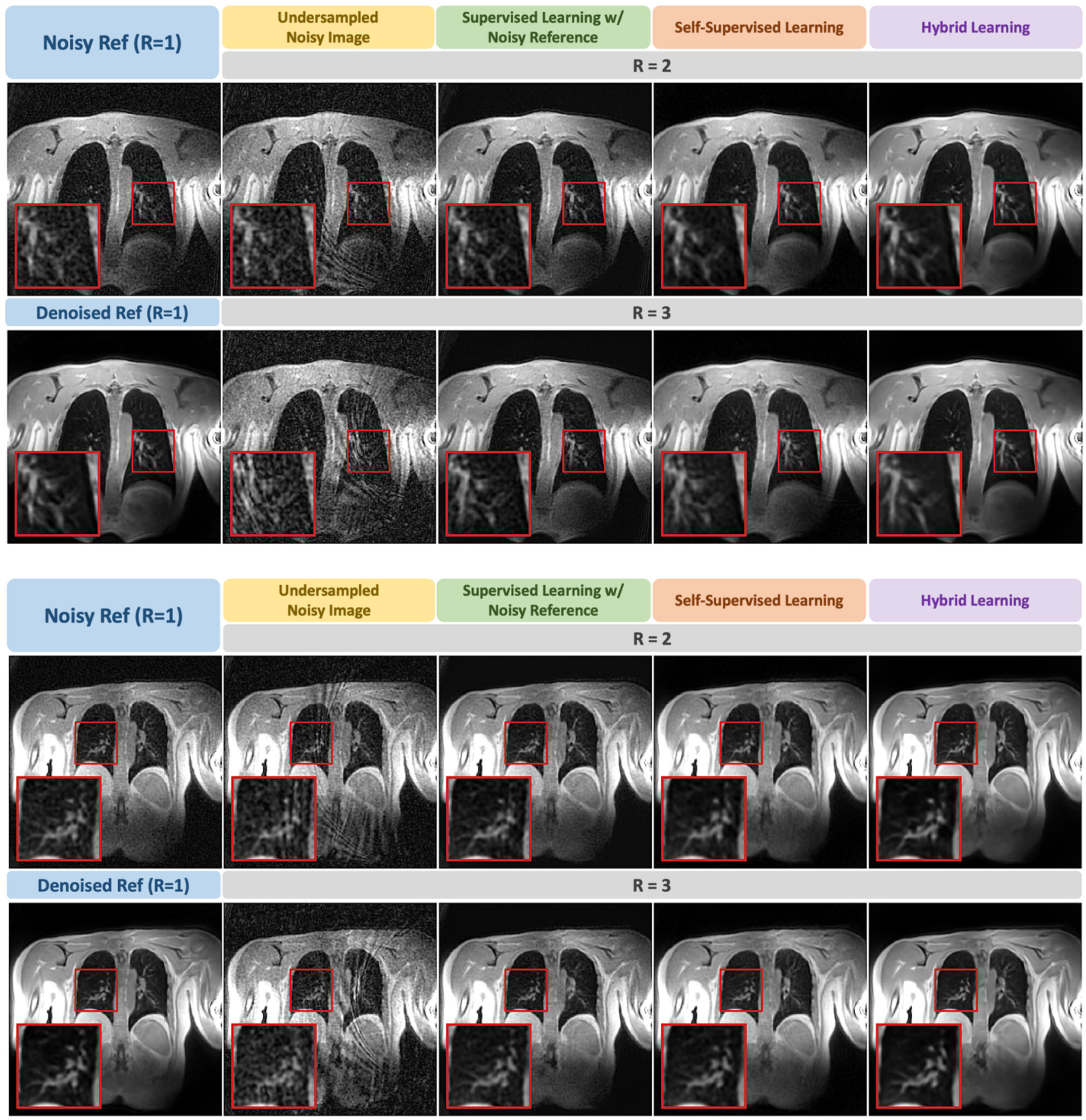

Two similar comparisons from the experiment on 0.55T lung MRI in two additional cases, where hybrid learning consistently yielded better image quality.